\documentclass[reprint, nofootinbib,  superscriptaddress]{revtex4-1}
\usepackage{amsmath} \usepackage{graphicx} \usepackage{dcolumn}
\usepackage{bm} 
\usepackage{amssymb}
\usepackage{pstricks}
\usepackage{subfigure}
\usepackage{siunitx} %for Angstrom units
\usepackage{mathtools}

\newrgbcolor{dred}{.67 0 0}

\begin{document}

\newlength{\figurewidth}
\setlength{\figurewidth}{0.6 \columnwidth}

\newcommand{\prtl}{\partial}
\newcommand{\la}{\left\langle}
\newcommand{\ra}{\right\rangle}
\newcommand{\dla}{\la \! \! \! \la}
\newcommand{\dra}{\ra \! \! \! \ra}
\newcommand{\we}{\widetilde}
\newcommand{\smfp}{{\mbox{\scriptsize mfp}}}
\newcommand{\smp}{{\mbox{\scriptsize mp}}}
\newcommand{\sph}{{\mbox{\scriptsize ph}}}
\newcommand{\sinhom}{{\mbox{\scriptsize inhom}}}
\newcommand{\sneigh}{{\mbox{\scriptsize neigh}}}
\newcommand{\srlxn}{{\mbox{\scriptsize rlxn}}}
\newcommand{\svibr}{{\mbox{\scriptsize vibr}}}
\newcommand{\smicro}{{\mbox{\scriptsize micro}}}
\newcommand{\scoll}{{\mbox{\scriptsize coll}}}
\newcommand{\sattr}{{\mbox{\scriptsize attr}}}
\newcommand{\sth}{{\mbox{\scriptsize th}}}
\newcommand{\sauto}{{\mbox{\scriptsize auto}}}
\newcommand{\seq}{{\mbox{\scriptsize eq}}}
\newcommand{\teq}{{\mbox{\tiny eq}}}
\newcommand{\sinn}{{\mbox{\scriptsize in}}}
\newcommand{\suni}{{\mbox{\scriptsize uni}}}
\newcommand{\tin}{{\mbox{\tiny (in)}}}
\newcommand{\tout}{{\mbox{\tiny (out)}}}
\newcommand{\scr}{{\mbox{\scriptsize cr}}}
\newcommand{\tstring}{{\mbox{\tiny string}}}
\newcommand{\sperc}{{\mbox{\scriptsize perc}}}
\newcommand{\tperc}{{\mbox{\tiny perc}}}
\newcommand{\sstring}{{\mbox{\scriptsize string}}}
\newcommand{\stheor}{{\mbox{\scriptsize theor}}}
\newcommand{\sGS}{{\mbox{\scriptsize GS}}}
\newcommand{\sBP}{{\mbox{\scriptsize BP}}}
\newcommand{\sNMT}{{\mbox{\scriptsize NMT}}}
\newcommand{\sbulk}{{\mbox{\scriptsize bulk}}}
\newcommand{\tbulk}{{\mbox{\tiny bulk}}}
\newcommand{\sXtal}{{\mbox{\scriptsize Xtal}}}
\newcommand{\sliq}{{\text{\tiny liq}}}

\newcommand{\smin}{\text{min}}
\newcommand{\smax}{\text{max}}

\newcommand{\saX}{\text{\tiny aX}}
\newcommand{\slaX}{\text{l,{\tiny aX}}}

\newcommand{\svap}{{\mbox{\scriptsize vap}}}
\newcommand{\sjam}{J}
\newcommand{\Tm}{T_m}
\newcommand{\sTS}{{\mbox{\scriptsize TS}}}
\newcommand{\sDW}{{\mbox{\tiny DW}}}
\newcommand{\cN}{{\cal N}}
\newcommand{\cB}{{\cal B}}
\newcommand{\br}{{\bf r}}
\newcommand{\be}{\bm e}
\newcommand{\cH}{{\cal H}}
\newcommand{\cHlt}{\cH_{\mbox{\scriptsize lat}}}
\newcommand{\sthermo}{{\mbox{\scriptsize thermo}}}

\newcommand{\bu}{\bm u}
\newcommand{\bk}{\bm k}
\newcommand{\bX}{\bm X}
\newcommand{\bY}{\bm Y}
\newcommand{\bA}{\bm A}
\newcommand{\bb}{\bm b}
\newcommand{\bx}{{\bm x}}
\newcommand{\by}{{\bm y}}
\newcommand{\bz}{{\bm z}}

\newcommand{\lintf}{l_\text{intf}}

\newcommand{\DV}{\delta V_{12}}
\newcommand{\sout}{{\mbox{\scriptsize out}}}
\newcommand{\dv}{\Delta v_{1 \infty}}
\newcommand{\dvin}{\Delta v_{2 \infty}}

\newcommand{\tp}{\widetilde{p}}
\newcommand{\tP}{\widetilde{P}}
\newcommand{\bK}{{\bm K}}
\newcommand{\bI}{{\bm 1}}
\newcommand{\tbP}{\widetilde{\bm P}}
\newcommand{\tps}{\widetilde{\psi}}
\newcommand{\bP}{{\bm P}}
\newcommand{\bp}{{\bm p}}
\newcommand{\bps}{{\bm \psi}}
\newcommand{\bpi}{{\bm \pi}}
\newcommand{\lth}{l_\text{th}}
\newcommand{\tMC}{{\mbox{\tiny MC}}}

\newcommand*\xbar[1]{%
  \hbox{%
    \vbox{%
      \hrule height 0.5pt % The actual bar
      \kern0.5ex%         % Distance between bar and symbol
      \hbox{%
        \kern-0.1em%      % Shortening on the left side
        \ensuremath{#1}%
        \kern-0.1em%      % Shortening on the right side
      }%
    }%
  }%
}

\newcommand{\cV}{{\cal V}}

\def\Xint#1{\mathchoice
   {\XXint\displaystyle\textstyle{#1}}%
   {\XXint\textstyle\scriptstyle{#1}}%
   {\XXint\scriptstyle\scriptscriptstyle{#1}}%
   {\XXint\scriptscriptstyle\scriptscriptstyle{#1}}%
   \!\int}
\def\XXint#1#2#3{{\setbox0=\hbox{$#1{#2#3}{\int}$}
     \vcenter{\hbox{$#2#3$}}\kern-.5\wd0}}
\def\ddashint{\Xint=}
\def\dashint{\Xint-}
\title{Emergence of pseudo-time during optimal Monte Carlo sampling
  \\ and temporal aspects of symmetry breaking and restoration}

\author{Yang He} \affiliation{Department of Chemistry, University of
  Houston, Houston, TX 77204-5003, USA}

\author{Vassiliy Lubchenko} \email{vas@uh.edu} \affiliation{Department
  of Chemistry, University of Houston, Houston, TX 77204-5003, USA}
\affiliation{Department of Physics, University of Houston, Houston, TX
  77204-5005, USA} \affiliation{Texas Center for Superconductivity,
  University of Houston, Houston, TX 77204-5002, USA}

\date{\today}

\begin{abstract}

  We argue that one can associate a pseudo-time with sequences of
  configurations generated in the course of classical Monte Carlo
  simulations for a single-minimum bound state, if the sampling is
  optimal. Hereby the sampling rates can be, under special
  circumstances, calibrated against the relaxation rate and frequency
  of motion of an actual physical system. The latter possibility is
  linked to the optimal sampling regime being a universal crossover
  separating two distinct suboptimal sampling regimes analogous to the
  physical phenomena of diffusion and effusion, respectively.  Bound
  states break symmetry; one may thus regard the pseudo-time as a
  quantity emerging together with the bound state. Conversely, when
  transport among distinct bound states takes place---thus restoring
  symmetry---a pseudo-time can no longer be defined. One can still
  quantify activation barriers, if the latter barriers are smooth, but
  the simulation becomes impractically slow and pertains to overdamped
  transport only.  Specially designed Monte Carlo moves that bypass
  activation barriers---so as to accelerate sampling of the
  thermodynamics---amount to effusive transport and lead to severe
  under-sampling of transition-state configurations that separate
  distinct bound states while destroying the said
  universality. Implications of the present findings for simulations
  of glassy liquids are discussed.

\end{abstract}

\maketitle

\section{Motivation}

The thermodynamics of a classical system can be efficiently quantified
by Gibbs-sampling its Boltzmann distribution~\cite{newman1999monte,
  doi:10.1063/1.471317, Berthier2023,
  doi:10.1063/1.436415} because the sampled variables are not subject
to inertia. Microscopic characterization of transition states for
activated escape from bound states, then, poses a challenge: To
quantify rates of escape, one must evaluate autocorrelations for
quantities sampled at well-defined intervals of {\em time}.  Yet it is
not clear to what extent sequences of configurations generated using
classical Monte Carlo (MC) simulation~\cite{newman1999monte}
correspond with actual dynamics, if at all. There is also the distinct
possibility that the transition states---which often contribute
negligibly little to thermodynamics---are not adequately represented
during statistical sampling thus preventing one from quantifying their
microscopic characteristics such as the extent of cooperativity.

At the same time, correlation functions generated in the course of
Monte Carlo sampling of the Boltzmann distribution for a classical
system often look qualitatively similar to correlation functions
measured in experiment. Thus one may reasonably
inquire~\cite{doi:10.1063/1.4902136, KIKUCHI1991335,
    Berthier_2007} whether there are conditions under which there may
be a quantitative connection between the apparent kinetics observed,
respectively, in Monte Carlo simulations and the actual molecular
dynamics. One may immediately object that already the lack of inertia
intrinsic to statistical sampling introduces too much ambiguity
because masses of individual particles and collective modes, if any,
are generally distributed. But the latter ambiguity may be only
semi-quantitative since a particle's proper time is scaled by the
square-root of the particle's mass, while in many systems of
interest~\cite{doi:10.1063/1.471317} the masses of what one would
ordinarily define as particles vary only modestly. For instance, a
methyl group weighs 15 Dalton, the amino group 17, the hydroxy group
17, etc.  In any event, the ambiguity due to mass variation, if any,
is often much smaller than the dynamic range of six orders of
magnitude---and usually more---that is found in many contexts of
practical interest,  such as dynamics of glassy
  liquids.~\cite{L_AP, LW_ARPC}
    
Accelerated Monte Carlo protocols---such as those employing peculiar
moves that swap particles' places in glassy
mixtures~\cite{PhysRevX.7.021039, Berthier2023, doi:10.1063/1.5086509,
  Berthier_2019}---seem to be particularly efficient at sampling the
thermodynamics very deep in the free energy landscape, where physical
moves would be subject to high activation barriers.~\cite{L_AP,
  LW_ARPC, PhysRevX.12.041028} Clearly, this efficiency is predicated
on the ability of an accelerated protocol to circumvent the transition
states for such activated processes. This, then, prompts one to
examine to what extent accelerated protocols sample those transition
states or might allow one to quantify their microscopic
characteristics. 

Here we argue that notwithstanding the apparent similarity of
MC-produced relaxation profiles to physical relaxations, one can {\em
  not} associate the time-step of Monte Carlo sampling with a physical
time in systems characterized by a distribution of length and time
scales. Such distribution is characteristic of most applications of
practical interest, of course.  Only under some special circumstances
can one define a pseudo-time for simulated sequences of variables so
that the time interval separating two consecutively sampled
configurations can be connected to the relaxation time and/or
vibrational period of a bound motion in an actual system.  That such
special circumstances could arise can be understood by combining the
following two notions:

On the one hand, configurations sampled during physical processes near
equilibrium are such that the rate of entropy increase is at its
maximum possible value, per Onsager.~\cite{PhysRev.37.405,
  PhysRev.38.2265} Thus the set of physical configurations
statistically relevant for processes near equilibrium also correspond
to the optimal {\em rate} of equilibration. At the same time,
equilibrium from the statistical viewpoint corresponds to the degrees
of freedom strictly obeying the Boltzmann distribution. Thus maximum
entropy production is equivalent to the rate of sampling of the
Boltzmann distribution being optimal.

On the other hand, the question of the optimal rate of Gibbs sampling
of probability distributions is well defined and, furthermore, can be
answered rather accurately in many cases of practical interest, see
Ref.~\cite{10.2307/3182776} and references therein. The optimal regime
of sampling emerges as a compromise between the length of attempted
increments of the sampled variable, on the one hand, and the
acceptance rate for attempted moves, on the other hand.  Suppose the
increment is Gaussianly distributed with standard deviation $l$ while
the Boltzmann distribution is a univariate Gaussian distribution with
standard deviation $\sigma$.  The optimal step
size~\cite{10.2307/3182776} is, apparently, rather large, $l \approx
2.4 \: \sigma$, suggesting the continuity of actual physical
trajectories may be statistically redundant under certain
circumstances. To see that such redundancy is plausible, consider an
equilibrated, mechanically-stable solid composed of hard spheres. One
can approximately associate the Monte Carlo step size with the
distance traveled between consecutive collisions and, thus, the
typical step size should be comparable to the double of the
r.m.s.d. for the vibrational motion of a given particle, within the
cage created by the surrounding particles. Motions between consecutive
collisions are straight lines that can be generically reconstructed,
since the typical speed is fixed by temperature.

Motivated by the notions above, one might further ask to what extent
Monte Carlo-generated correlation functions could also reflect the
kinetics of activated transitions among distinct bound states, and to
what extent the sampled configurations could reflect the morphology of
the corresponding transition states. This is a question of
considerable interest in systems exhibiting free energy surfaces with
multiple minima, such as glassy liquids.~\cite{L_AP} In the presence
of activated processes, relaxation functions show two distinct,
time-separated processes:~\cite{MezeiRussina} The faster process
presumably corresponds to vibrational relaxation within individual
free energy minima, while the slower process corresponds to activated
escapes from those free-energy minima.~\cite{L_AP, LW_ARPC, LW_Wiley}
If one could connect Monte Carlo sampling of individual bound states
to the actual vibrational dynamics, one could perhaps use the
corresponding sampling rates as a reference timescale to
systematically quantify the lifetimes of bound states.  It would be of
value, too, if one could calibrate activated kinetics---by using
vibrational relaxation rates as the reference---to compare results of
simulations across systems with different force fields.

In addressing these questions, we proceed as follows: In
Section~\ref{FP}, we first construct an explicit semi-Markov process
corresponding to the Metropolis variety of Monte-Carlo sampling, which
allows one to associate a continuous time-like variable to the
intrinsically discrete process of statistical sampling. We show that
the two extreme limits of a very small and very large size of
attempted displacements are analogous to physical phenomena of
diffusion and effusion, respectively.  At the crossover between the
two regimes, the sampling rate or, equivalently, the relaxation rate
reaches its optimal value. At the same time, universal relationships
emerge between the time interval of the simulation---which is \`{a}
priori an arbitrary parameter---and the apparent relaxation rate, as
well as the frequency of motion within a single-well bound state. We
then outline the special circumstances under which the aforementioned
time variable of the semi-Markov process can be thought of as a
pseudo-physical time.  The crossover is also linked to the appearance
of a well-defined gap separating the two largest eigenvalues of the
transition matrix for the Markov chain corresponding to the Monte
Carlo sampling. In Section~\ref{contdiscr}, we show that
notwithstanding the apparent causal connection among the
configurations generated during optimal sampling, the latter
configurations cannot be interpreted as snapshots of continuous
trajectories, thus greatly limiting one's ability to sample
transitions states separating distinct bound states. There we also
discuss ambiguities, inherent to Monte Carlo simulations, that stem
from the distribution of particles' masses. The above findings are
explicitly illustrated in Section~\ref{examples} using both analytical
estimates and direct simulation. There, we show that in the
  presence of activated transport, a pseudo-time cannot be defined.
In the strict diffusion limit and for sufficiently smooth barrier
tops, one can still hope to recover correct activation energies. The
diffusion limit---in which the trajectories become effectively
continuous, though non-inertial---is however computationally
inefficient. Moves designed to speed up sampling {\em also} lead to a
loss of the connection between optimal sampling of individual bound
states and vibrational relaxation. Furthermore, such
  accelerated protocols result in severe under-sampling of transition
  states for activated transport and underestimation of activation
  barriers. Finally we discuss implications of the present results
for the ongoing efforts to elucidate the detailed mechanism of
activated transport in glassy liquids.

%\newpage

\section{Monte Carlo simulation as a semi-Markov process, and emergence
of optimal sampling}
\label{FP}
  
First we define a Markov chain with stationary transition
probabilities~\cite{Ross} whose sole purpose is to mutually connect a
set of physical states of interest, irrespective of the long-term
probability to reside in any one of those states.  Thus we introduce a
transition matrix that specifies the probability to be in state $j$ at
step number $N+1$, if the system was in state $i$ at step number $N$,
irrespective of the prior history:
\begin{equation}
  q_{ji} \equiv q(j \leftarrow i).
\end{equation}
In the context of Monte Carlo (MC) simulation, the matrix $q_{ji}$
specifies the set of trial moves that could be attempted, in
principle, during sampling and is often called the ``proposal
density.''  We are exclusively interested in probability conserving
processes. That is,
\begin{equation} \label{qcons} \sum_j q_{ji} = 1,
\end{equation}
for all $i$'s. Note the above conventions for the order of the matrix
indices are the opposite of what is often adopted in
statistics. Within the present conventions, all operators consistently
apply to the right, as would the conventional derivative, for
instance.

One may further ask: Can one modify the proposals specified by the
matrix $q_{ji}$ so that the resulting transition matrix $\pi_{ji}$
yields a pre-specified stationary probability distribution of choice:
$p_i$, $\sum_i p_i = 1$? The sought transition matrix must obey
detailed balance:
\begin{equation} \label{detailedbal} \pi_{ji} p_i = \pi_{ij} p_j.
\end{equation}
Per Hastings,~\cite{10.1093/biomet/57.1.97} there are an infinite
number of prescriptions to accomplish this task. Hereby one introduces
a new quantity $\alpha_{ji}$, often called the ``acceptance rate'',
that multiplies the off-diagonal elements of the proposal matrix to
yield the off-diagonal elements of the sought matrix $\pi_{ji}$:
\begin{equation} \label{detbal} \pi_{ji} =  \alpha_{ji}  q_{ji},
    j \ne i 
\end{equation}
while setting the diagonal element so as to ensure probability
conservation:
\begin{equation}  \label{picons} \pi_{ii} =  1 - \sum_{j \ne i} \pi_{ji}
\end{equation}
The following prescription for the acceptance rate:
\begin{equation} \label{accrate} \alpha_{ji} = \left\{ \begin{array}{ll} 1, &
    \frac{q_{ij} p_j}{q_{ji} p_i} \ge 1 \\ \\ \frac{q_{ij} p_j}{q_{ji}
      p_i}, & \frac{q_{ij} p_j}{q_{ji} p_i} < 1
  \end{array}
  \right.
\end{equation}
together with a symmetric proposal density:
\begin{equation} \label{qsym} q_{ij} = q_{ji} 
\end{equation}
yield the venerable Metropolis selection
criterion:~\cite{doi:10.1063/1.1699114}
\begin{equation}  \label{metropolis} \alpha_{ji} = \left\{ \begin{array}{ll} 1, &
    \frac{p_j}{p_i} \ge 1 \\ \\ \frac{p_j}{p_i}, & \frac{p_j}{p_i} < 1
  \end{array}   \right.
\end{equation}
One may parametrize the probability distribution $p_i$, without loss
of generality, as a Boltzmann distribution using an energy-like
parameter $E_i$:
\begin{equation} p_i \propto e^{-\beta E_i}  \equiv e^{-E_i/k_B T}
\end{equation}
where $T \equiv 1/k_B \beta$ stands for temperature.  For the sake of
completeness we note that bound states generally have both enthalpic
and entropic contributions. Consequently, the appropriate energy $E_i$
could include the isobaric contribution $p V_i$ and/or grand-canonical
contribution $(-\mu N_i)$, where $p$ stands for pressure, $V$ volume,
$\mu$ the chemical potential, and $N$ particle number.

It will be convenient to introduce an auxiliary function
\begin{equation} \label{Bji} B(j \leftarrow i) \equiv
  \left\{ \begin{array}{ll} 1, & E_j \le E_i \\ \\ e^{-\beta (E_j -
      E_i)}, & E_j > E_i
  \end{array}   \right.
\end{equation}
which is simply the acceptance ratio for the Metropolis algorithm
complemented by the value $B(i \leftarrow i)=1$; the function $B(j
\leftarrow i)$ is thus continuous with respect to both
arguments. Using an arrow in the notation is unconventional yet it
seems to make algebraic manipulations more vivid.

Eqs.~(\ref{detbal})-(\ref{Bji}) can be consolidated to yield the
following expression for the sought transition matrix:
\begin{align}  \label{piji1} \pi_{ji} & = q_{ij} B(j \leftarrow i),
  \hspace{3mm} (j \ne i) \\ \pi_{jj} & = q_{jj} B(j \leftarrow j) +
  \sum_k q_{jk} [1 - B(k \leftarrow j)] \nonumber \\ & = 1 + q_{jj}
  B(j \leftarrow j) - \sum_k q_{jk} B(k \leftarrow j),
  \label{piji2} 
\end{align}
where we used Eqs.~(\ref{qcons}) and (\ref{qsym}). Note the summation
over $k$ in Eq.~(\ref{piji2}) is unrestricted.

One may further associate a semi-Markov process~\cite{Ross} to the
transition matrix $\pi_{ji}$ by specifying a distribution
$\psi_{ji}(t)$ of wait (or, interarrival) times $t$ for each of the $i
\to j$ transitions; the latter distributions must each be normalized
$\int dt \, \psi_{ji}(t) = 1$.  One may model an arbitrary process by
specifying an appropriate set of functions $\psi_{ji}(t)$. One thus
obtains the probability $P_{ji}(t)$ of being in state $j$ at time $t$
given the particle was in state $i$ at time zero. At any time, $\sum_j
P_{ji}(t) = 1$. Although we will often refer to the time-variable $t$
of the semi-Markov process simply as ``time,'' for brevity, it is
unrelated to physical time except under special circumstances, see
below.

If one specifies such initial conditions that at time zero:
$P_{ij}(0)=\delta_{ij}$, the matrix $\bP(t) \equiv \{P_{ij}(t)\}$ can
be thought of as a Green's function. Indeed, the latter matrix yields
the population pattern in the system at time $t$: $\bp(t) = \bP(t)
\bp(0)$, where $\bp(t) \equiv [p_1(t) \:\: p_2(t) \ldots ]^T$ is a
column whose entries are state populations $p_i(t)$ at time $t$.  The
Laplace transform of the probability, $\tbP(s) = \int_0^\infty dt \,
e^{-st} \bP(t)$, can be expressed in terms of the transition-matrix
elements $\pi_{ij}$ and the Laplace transforms of the functions
$\psi_{ij}$, through a well known formula,\cite{Ross} see also
Ref.~\cite{LScontrol}:
\begin{equation}
\tbP(s)^{-1}|_{ij} = [\delta_{ij} - \pi_{ij} \tps_{ij}(s)]
\frac{s}{1-\sum_k \pi_{kj} \tps_{kj}(s)}.
\label{Psu}
\end{equation}

The formal expression (\ref{Psu}) is generally intractable but
simplifies greatly when the waiting time distribution depends only on
the identity of the originating site: $\psi_{ji}(t) =
\psi_i(t)$. Here, we are interested in the simpler yet, fully uniform
case
\begin{equation} \label{psitu} \psi_{ij}(t) = \psi(t).
\end{equation}
Hereby we assign a distributed, temporal-like interval to each Monte
Carlo step where the assignment is independent of the current
configuration of the system.  Under the constraint from
Eq.~(\ref{psitu}), Eq.~(\ref{Psu}) yields:
\begin{equation}
s \tbP(s) -  \bI = \frac{s \tps(s)}{1-\tps(s)} (\bpi-\bI)  \tbP(s)
\label{Psuuni}
\end{equation}
where $\bpi$ is the shorthand for the square matrix composed of
elements $\pi_{ij}$ and $\bI$ is the unit matrix.

In conventional Monte Carlo simulations, one simply uses the step
number as the effective time variable:
\begin{equation} \label{psiMC} \psi_\tMC (t) = \delta(t - \Delta t)
\end{equation}
where $\Delta t = 1$. We will adhere to this convention when
performing Monte Carlo simulations as well. For calculations, on the
other hand, it is more convenient to make the time $t$ of the
semi-Markov process a continous variable, in a statistical sense, by
adopting a smooth probability density that is non-vanishing at the
origin. To this end, we will adopt the Poisson statistics:
\begin{equation} \label{pPoisson} \psi(t) = \frac{1}{\Delta t} e^{-t/\Delta t}.
\end{equation}
The parameter $\Delta t$ is now seen to specify the {\em average}
waiting time for the semi-Markov process. Eq.~(\ref{Psuuni})
immediately gives a first-order differential equation for the survival
probabilities, in terms of a continuous, time-like variable $t$:
\begin{equation}
\dot{\bP}(t) = \frac{1}{\Delta t} (\bpi-\bI) \,  \bP(t)
\label{Ptuni}
\end{equation}
and the initial condition $\bP(t=0)=\bI$.   Equivalently, one may
directly write down a differential equation for a probability
distribution
\begin{equation}
\dot{\bp}(t) = \frac{1}{\Delta t} (\bpi-\bI) \, \bp(t)
\label{Ptuni1}
\end{equation}
where one must specify the initial condition $\bp(t=0)$.

Within the assumptions that led to Eq.~(\ref{Ptuni1}), the question of
the apparent rate of Monte Carlo sampling is reduced to the question
of the eigenvalues of the matrix $\bpi$. The largest eigenvalue,
$\lambda_1=1$, has the equilibrium probability distribution as its
eigenvector, per Eqs.~(\ref{detailedbal}) and (\ref{picons}). The
corresponding relaxation rate vanishes: $-(\lambda_1-1)=(1-\lambda_1)
= 0$. Consider now the next largest eigenvalue, $\lambda_2$. The
quantity $(1-\lambda_2)$ yields the lower bound on the rate at which
the apparent distribution approaches its equilibrium value, according
to Eq.~(\ref{Ptuni1}). As such, the quantity $(1-\lambda_2)$ also
measures the efficiency of sampling. This is also a good place to
elaborate on the effect of a specific choice of the functional form
for the wait-time distribution $\psi$.  Suppose we use an eigenvector
corresponding to some eigenvalue $\lambda$ of the matrix $\bpi$ as the
initial condition for the semi-Markov process. For the discrete
prescription from Eq.~(\ref{psiMC}), the probability value after $N$
events is $p(N = t/\Delta t) = \lambda^N p(t=0) = e^{t \ln
  \lambda/\Delta t} p(t=0)$, which nominally corresponds to the rate
$(-\ln \lambda/\Delta t)$. On the other hand, the Poisson prescription
from Eq.~(\ref{pPoisson}) gives $p(t) = e^{-(1-\lambda) t/\Delta t}
p(t=0)$, per Eq.~(\ref{Ptuni1}), which corresponds to the rate
$(1-\lambda)/\Delta t$. The quantities $(-\ln \lambda)$ and
$(1-\lambda)$ both monotonically decrease with $\lambda$, and so the
difference between these two specific prescriptions for the
distribution $\psi$ of the wait times does not affect the discussion
of the efficiency of sampling in a substantive way. In any event, the
rates $(-\ln \lambda)$ and $(1-\lambda)$ approach each other
asymptotically in the $\lambda \to 1$ limit and are numerically close
even in the worst-case scenario, as we shall see below.  Other forms
for the wait-time distribution $\psi$ could be considered, in
principle. For the reader's reference, we show in
Appendix~\ref{psitspec} that the Laplace transform of the relaxation
profile for eigenvectors corresponding to an eigenvalue $\lambda$, for
an arbitrary $\psi(t)$, is given by a rather simple equation:
\begin{equation} \label{plambdat}
  \tp_\lambda(s) = \frac{1}{s} \: \frac{1-\tps(s)}{1 - \lambda \tps(s)}
\end{equation}
Although the resulting relaxation profile $p_\lambda(t)$ can be quite
complicated---depending on the specific form of $\tps(s)$---we see it
is always slower for higher values of $\lambda$. In any event, it
seems prudent to avoid $\psi$'s with long tails unless slow processes
are known to be present.

Next we ask whether there are circumstances under which the time-like
quantity $t$ may be connected to actual physical time.  To this end,
we explicitly rewrite Eq.~(\ref{Ptuni}) for a process in which
individual sites are each associated with a location in space of
dimensionality $d$. The latter dimensionality may or may not be equal
to the number of physical dimensions. And so, for instance, if one
attempts to move two particles at a time, in the actual physical
space, the dimensionality of the semi-Markov process is six.  Going
over to continuous variables, $\bP_{ij} \to P(\bx, \bz)$ etc.,
Eqs.~(\ref{piji1})-(\ref{piji2}) and (\ref{Ptuni}) straightforwardly
yield:
\begin{align} \dot{P}(\by, \bz) & = \frac{1}{\Delta t} \int d^d \bx
  \: q(\bx, \by) \nonumber \\ & \times [B(\by \leftarrow \bx) P(\bx,
    \bz) - B(\bx \leftarrow \by) P(\by, \bz)]
\end{align}
and initial condition $P(\by, \bz)_{t=0}= \delta^{(d)}(\by-\bz)$,
where $\delta^{(d)}$ is the $d$-dimensional Dirac delta-function. The
above equation can be equivalently rewritten as an
integro-differential equation for a function of a single set of
coordinates:
\begin{equation} \label{MEcont} \dot{p}(\by) =
  \frac{1}{\Delta t} \int \! d^d \bx \hspace{.5 mm} q(\bx, \by) [B(\by \leftarrow
    \bx) p(\bx) - B(\bx \leftarrow \by) p(\by)]
\end{equation}
where the initial condition $p(\by, t=0)$ is user-specified. To
emphasize that the contribution of the kinetic energy to the total
energy is not included in classical Monte Carlo sampling, from here on
we switch notations as follows:
\begin{equation} E_i \to V(x)
\end{equation}
It may be sometimes more convenient to rewrite Eq.~(\ref{MEcont}) for
an auxiliary function defined as
\begin{equation} \tp (\bx) \equiv p(\bx) e^{\beta V(\bx)}
\end{equation}
whereby deviations from equilibrium, if any, result in the function
$\tilde{p}$ being spatially non-uniform. One thus obtains
straightforwardly:
\begin{align} \label{contmod} (\Delta t) \: \dot{\tp}_\bx & =
   \int_{V_\by < V_\bx} d^d \by \: (\tp_\by - \tp_\bx) q(\bx, \by)
   \nonumber \\ & + \int_{V_\by > V_\bx} d^d \by \: (\tp_\by -
   \tp_\bx) e^{-\beta(V_\by-V_\bx)} q(\bx, \by)
\end{align}
and we have compactified notations for typographical clarity: $p_x
\equiv p(\bx)$, $V_\bx \equiv V(\bx)$, etc.

From here on, we adopt the translationally invariant, Gaussian
proposal density:
\begin{equation} \label{propdG} q(\bx, \by) =
  \frac{1}{(2 \pi l^2)^{d/2}} e^{-\frac{(\bx-\by)^2}{2 l^2}}
\end{equation}
where the quantity $l$ thus gives the r.m.s.d. of the distribution of
the step size for attempted moves along a single spatial dimension.

In the ultra-local limit of a vanishing step size, one may
Taylor-expand the function $\tp_y$ in the integrands in
Eq.~(\ref{contmod}) around $y=x$, which yields straightforwardly:
\begin{equation} \dot{\tp}  = \frac{l^2}{2 \Delta t}
  \left[\nabla^2 \tp - (\nabla \beta V) (\nabla \tp) \right].
\end{equation}
Subsequently, this leads to the familiar Smoluchowski-Fokker-Planck
equation~\cite{zwanzig2001nonequilibrium, vanKampen} for the original
probability density:
\begin{equation} \dot{p}  = D_\tMC
  \nabla e^{-\beta V} \nabla e^{\beta V} p 
\end{equation}
after we formally associate  an effective diffusivity
\begin{equation} \label{Ddef} D_\tMC \equiv \frac{l^2}{2 \Delta t}
\end{equation}
with the step size $l$ and the parameter $\Delta t$, consistent with
Ref.~\cite{KIKUCHI1991335} We use the label ``MC'' to emphasize that
the quantity $D_\tMC$ is not a physical diffusivity. The Smoluchowski
equation above formally corresponds to an overdamped Langevin
dynamics,~\cite{Goldenfeld}
\begin{equation} \label{Langevindamp} \zeta_\tMC \, \dot{\bx} =
  - \frac{\prtl V}{\prtl \bx} + {\bm f}_\text{th}
\end{equation}
Where the quantity $\zeta_\tMC$ is an effective friction coefficient
defined through the diffusivity (\ref{Ddef}) via an Einstein-like
relation:
\begin{equation} \label{zeta} \zeta_\tMC = \frac{k_B T}{D_\tMC}
\end{equation}
and ${\bm f}_\text{th}$ is the corresponding, effective fluctuating
force needed for energy balance. For a $d$-dimensional oscillator:
\begin{equation} \label{EHO} V(\bx) = \frac{k x^2}{2}
\end{equation}
the overdamped Langevin dynamics yields, by Eq.~(\ref{Langevindamp}),
the following relaxation rate:
\begin{equation} \label{taullshort} \frac{1}{\tau} = \frac{k}{\zeta_\tMC}
  = \frac{1}{2 \Delta t} \left(\frac{l}{\lth} \right)^2
\end{equation}
where we used Eqs.~(\ref{Ddef}) and (\ref{zeta}) and have introduced
the thermal length $\lth$ that determines the r.m.s.d. for bound
motions along an individual spatial direction, per the Boltzmann
distribution corresponding to the energy function (\ref{EHO}):
\begin{equation} \label{lthdef} \lth \equiv
  \left( \frac{k_B T}{k} \right)^{1/2}.
\end{equation}

When the step size $l$ becomes comparable to or greater than the
length scale for spatial variation of the probability distribution,
the long-wavelength expansion used to derive the Fokker-Planck
equation is no longer valid, while the semi-Markov process
corresponding to MC sampling can not be thought of as diffusion-like
but, instead, becomes expressly non-local.  While the integral
equation (\ref{MEcont}) seems difficult to solve under general
circumstances, the limit of large $l$, $l \gg \lth$, can be readily
understood at a semi-quantitative level and, further, linked to the
physical phenomenon of {\em effusion}. Indeed, in this limit, the
region spanned by the equilibrium probability distribution is much
smaller than the region typically traversed in one jump. For energy
function (\ref{EHO}), the acceptance rate for jumps out of a typical
configuration, $\simeq \frac{l^d - \lth^d}{l^d} \, e^{-l^2/2 \lth^2}
\approx e^{-l^2/2 \lth^2}$, is very low, implying the system will
remain within bounds prescribed by the Boltzmann distribution most of
the time, while the non-typical configurations will be populated
roughly at $e^{-l^2/2 \lth^2}$. It is instructive to group all the
typical states into one set, call it A, while grouping the non-typical
states into one set also, call the latter set B.  Once in state B, the
system is at distance $l$ or so from the typical location and will
typically sample a region of volume $l^d$, up to a geometric factor,
since all of this region corresponds to acceptance rates comparable to
1. Only a fraction $\approx (\lth/l)^d$ of these moves will bring the
system back to state A, however, because the volume of the
corresponding region is $\lth^d$, up to a geometric factor. This
fraction, then, sets the sampling rate for the multivariate Gaussian
distribution corresponding to the bound state described by
Eq.~(\ref{EHO}), roughly at
\begin{equation} \label{taulllong} \frac{1}{\tau} \sim
   \frac{1}{\Delta t} \left( \frac{\lth}{l} \right)^d.
\end{equation}
where $l \gg \lth$.  Thus the probability to ``hit the spot'' occupied
by typical states is determined by the volume $\lth^d$ of the latter
region in the same way the chance for a particle to effusively escape
from a container is determined by the probability to hit the orifice,
hence the effusion analogy.

A somewhat more systematic approach, see Appendix~\ref{effusion},
confirms the above scaling while also yielding a geometric factor that
depends on the dimensionality of space:
\begin{equation} \label{taulllongdMT} \frac{1}{\tau} \approx
  \frac{1}{\Delta t} \left[ \frac{2}{d} \:
    \frac{\Gamma(d)}{\Gamma^2(d/2)} + 1 \right] \left( \frac{\lth}{l}
  \right)^d.
\end{equation}

\begin{figure}
  \includegraphics[width=  \columnwidth]{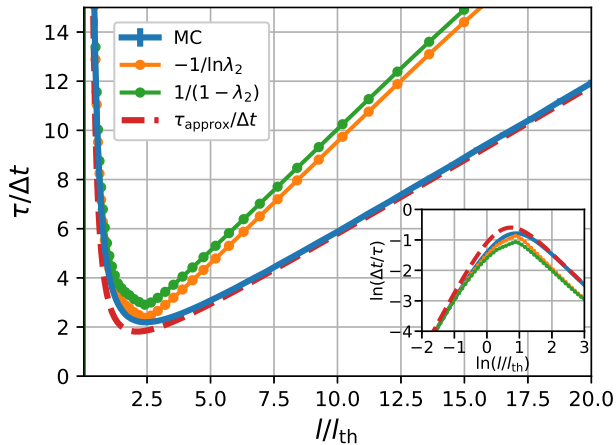}
  \caption{\label{llthFig} Thick solid line (``MC''): The longest
    relaxation time of the Monte-Carlo sampling of a harmonic bound
    state as a function of the step size relative to the thermal
    length, $l/\lth$. Thick dashed line (``$\tau_\text{approx}/\Delta
    t$''): $-1/\ln(1-\Delta t/\tau)$, where the $\Delta t/\tau$ ratio
    is computed according to the approximate relation (\ref{taull}) at
    $d=1$. Thin lines with dots show the result of direct
    diagonalization of the transition matrix of the Markov chain. We
    provide values for both the quantity $(1-\lambda_2)$ and $-\ln
    \lambda_2$. The results of diagonalization become progressively
    inaccurate at large values of $l$ because of numerical
    issues. Inset: Log-log plots of the corresponding relaxation rates
    vs. $l/\lth$.}
\end{figure}

Thus the efficiency of sampling declines with the step size $l$ for
large values of $(l/\lth)$, the rate of the decline increasing with
the dimensionality of the sampled variable.  Combined with the low-$l$
scaling from Eq.~(\ref{taullshort}), this means the sampling rate is
maximized for some value of the $l/\lth$ ratio intermediate between
the two asymptotic regimes described by Eqs.~(\ref{taullshort}) and
(\ref{taulllongdMT}), respectively. The position of the crossover,
then, fixes the dimensionless quantities $l/\lth$ and $\Delta t/\tau$
at some  values that can depend only on the dimensionality $d$ of space
and, thus, are universal. One can make an informal argument for the
$d$-dependence of the optimal displacement by noting that a particle
``moving'' according to the proposal density (\ref{propdG}) in $d$
dimensions tends to drift {\em away} from a given locale typically at
the rate of $l \sqrt{d}$ per step. At optimal sampling, a single move
should be able to traverse the thermally occupied region, whose size
is about $2 \lth$ across, irrespective of dimensionality. Thus we
conclude that, roughly, $l_\text{opt} \simeq 2 \lth/\sqrt{d}$ and that
optimal sampling becomes increasingly diffusion-like in higher
dimensions because the step size becomes progressively smaller than
the length $\lth$ characterizing the spatial extent of the Boltzmann
distribution.

The existence of a most efficient sampling rate had been established
systematically quite a while ago,\cite{10.2307/3182776} of course.  In
addition to establishing the optimal step size $l_\text{opt} \approx
2.4 \: \lth/\sqrt{d}$, alluded to in the Introduction, that prior work
also indicates that the optimal {\em rate} scales inversely
proportionally with the dimensionality $d$.

In Appendix~\ref{effusion} we provide an approximate, closed-form
expression, Eq.~(\ref{taull}), for the sampling rate as a function of
both the $l/\lth$ ratio and space dimensionality. This expression
interpolates between the short- and long-$l$ asymptotics for the
sampling rate $\Delta t/\tau$, Eqs.~(\ref{taullshort}) and
(\ref{taulllongdMT}) respectively, while approximately conforming to
the aforementioned asymptotics for the optimal step size and sampling
rate reported in Ref.~\cite{10.2307/3182776} and references
therein. Specifically we obtain,
\begin{equation}  \label{lopt} \frac{l_\text{opt}}{\lth} =
  \frac{C_1 (d)}{d^{1/2}}
\end{equation}
and
\begin{equation}  \label{tauopt} \frac{\Delta t}{\tau_\text{opt}} =
  \frac{C_2(d)}{d}.
\end{equation}
where the $d$-dependences on the r.h.s. of the two equations are
presented so as to highlight the large-$d$ scaling of the overall
expressions.  The quantities $C_1(d)$ and $C_2(d)$ are slow-varying
functions of $d$; both are numerically of order one and tend to steady
values as $d \to \infty$, see Appendix~\ref{effusion}. The numerical
value of the aforementioned interpolative expression,
Eq.~(\ref{taull}), is shown with the thick dashed line in
Fig.~\ref{llthFig} and agrees well with the apparent relaxation rate
obtained in a Monte Carlo simulation, shown as the thick solid line.
We determine the apparent MC relaxation rate $1/\tau$ by fitting the
two-point correlation function $\la x(0) x(t) \ra$ with the function
$\text{const} \times e^{-t/\tau}$. For proper comparison of the
closed-form expression (\ref{taull}) for the $\tau/\Delta t$ ratio
with MC data, we show the quantity $-1/\ln(1-\Delta t/\tau)$ and set
$\Delta t=1$, per the discussion preceding Eq.~(\ref{plambdat}).

One may further ask if sequences of configurations generated in the
course of optimal Monte Carlo sampling of a bound state can be put in
correspondence with a physical relaxation process. The answer is yes,
but under rather special circumstances: Suppose for the sake of
argument that the Monte Carlo moves approximate well the motions of
actual particles.  For example, imagine a mechanically stable solid
made of hard spheres. Particles move freely between consecutive
collisions. Within Einstein's approximation,~\cite{Einstein,
  doi:10.1080/00268978400101071, dens_F1} the displacement between
consecutive collisions is determined by the one-particle density
distribution function. The latter distribution is Gaussian for a
general harmonic solid;~\cite{RL_Tcr} anharmonic and many-body effects
amount to a correction. The proposal density (\ref{propdG}), too, is
Gaussian. Suppose further that one has determined, by trial and error,
the {\em optimal} value of step size $l$ for the simulation. At the
same time, Onsager had shown that typical configurations must maximize
the rate of equilibration~\cite{PhysRev.37.405, PhysRev.38.2265} and,
hence, optimize the rate of sampling of the Boltzmann
distribution. Thus under the special circumstances described in this
paragraph, one may approximately associate one-particle Monte Carlo
moves with the mean free path, i.e. the typical displacement between
two consecutive collisions.  The corresponding time interval must be
associated with the velocity auto-correlation time, i.e., the typical
timescale on which a certain fixed fraction of particles will have
been scattered. Thus at optimal sampling:
\begin{align}  \label{llconstr}
  l_\text{opt} & \approx l_\smfp \\ \tau_\text{opt} & \approx
  \tau_\sauto \label{ttconstr}
\end{align}
Consequently, Eqs.~(\ref{tauopt}) and (\ref{ttconstr}), when combined,
allow one to assign an actual value to the time-unit $\Delta t$ of the
semi-Markov process for MC sampling, which is otherwise an entirely
arbitrary quantity not linked to any physical phenomenon.

Scattering implies a change in the velocity, which is subject to
inertia. One may impose effects of inertia by adding the acceleration
term $m \ddot{x}$ to the equation of motion
(\ref{Langevindamp}).~\cite{zwanzig2001nonequilibrium, vanKampen} In
the Rayleigh limit, this corresponds to an auto-correlation time
\begin{equation} \tau_\sauto = \frac{m}{\zeta}
\end{equation}
while the corresponding diffusivity is given by
\begin{equation} \label{Dphys} D = \frac{l_\smfp^2}{2 \tau_\sauto},
\end{equation}
and $l_\smfp$ is the mean-free path, by construction.
c.f. Eq.~(\ref{Ddef}).  Combining these with Einstein's relation
$\zeta = k_B T/D$ and energy equipartition, $m v_\text{th}^2 \equiv m
\la v^2 \ra = k_B T$, one obtains:
\begin{equation} \label{kinematic} l_\smfp = \sqrt{2} v_\text{th}
    \tau_\sauto = \sqrt{2} (k_B T/m)^{1/2} \tau_\sauto.
\end{equation}

Using Eqs.~(\ref{lopt}), (\ref{ttconstr}), and (\ref{kinematic}) and
expressing the spring constant through the oscillation frequency
$\omega$, $k=m \omega^2$, we obtain that the optimal value of the MC
relaxation time is connected to the frequency of motion within the
bound state in a universal fashion:
\begin{equation}  \label{timemerged}  \tau_\text{opt}
  \simeq \frac{C_1(d)}{\sqrt{2 d}} \: \frac{1}{\omega}  \xrightarrow[d \to
    \infty]{} \frac{2}{\sqrt{d}} \: \frac{1}{\omega} 
\end{equation}
At $d=1$, $C_1(d)/\sqrt{2 d} \approx 1.5$.  Equation above connects
two physically observed quantities, by virtue of Eq.~(\ref{ttconstr}),
and thus serves as an internal check of the validity of the assumption
underlying Eqs.~(\ref{llconstr}) and (\ref{ttconstr}). If indeed
valid, the relation in Eq.~(\ref{timemerged}) implies the time
interval $\Delta t$ of the Monte Carlo simulation can be connected
with the frequency $\omega$  through a universal
relation:
\begin{equation} \label{tomega}  \Delta t =
  \frac{C_1(d) C_2(d)}{\sqrt{2} d^{3/2}} \: \frac{1}{\omega}
  \xrightarrow[d \to \infty]{} \frac{8/e}{d^{3/2}} \: \frac{1}{\omega}
\end{equation}
where we used Eq.~(\ref{tauopt}).  At $d=1$, $C_1(d) C_2(d)/\sqrt{2}
d^{3/2} \approx 0.63$.

The above example of a solid made of hard particles is of relevance to
a set of popular models that have been used to study glassy liquids,
see Section~\ref{conclusions}; yet it represents a specialized setup
and does not apply to those many contexts where the interaction range
is greater than the distance between the repulsive cores of the
neighboring particles. The approximation of moves being one-particle
becomes increasingly poorer with an increasing interaction
range. Hereby a progressively greater number of variables become
explicitly involved in evaluating the energy change resulting from
moving already a single particle. The question of the detailed form of
sampling moves that approximate the actual motions, as well as the
moves' optimal magnitude, remains well defined, but becomes difficult
to answer. This notion is consistent with findings by Berthier and
Kob~\cite{Berthier_2007}, who have observed that the vibrational wing
of MC-produced relaxation profiles shows a substantially broader
distribution of apparent relaxation times than that obtained using
Newtonian dynamics. In any event, because of the increased
dimensionality of the move, the optimal value of displacement for an
individual particle will decrease.  This will make the sampling more
diffusive in character and, hence, less efficient, per
Eqs.~(\ref{lopt}) and (\ref{tauopt}).

One may recall that the physical effect of friction can {\em also} be
thought of as arising from multiple, frequent interactions with the
environment that are each individually-weak. Thus, in order for
low-dimensional MC moves to be able to approximate actual motions
within bound states, the latter actual motions should not be overly
dampened, the formal criterion for internal consistency given by
Eq.~(\ref{timemerged}). Incidentally, Eq.~(\ref{timemerged}) in low
dimensions, $d \simeq 10^0$, happens to essentially coincide with the
condition for critical damping for the empirical damped-oscillator
model, whose relaxation dynamics is described by the equation
$\ddot{x} + \tau^{-1}_\sauto \dot{x} + \omega^2 x = 0$, per
Eq.~(\ref{ttconstr}). This apparent connection between the
diffusive-effusive and overdamped-underdamped crossovers,
respectively, is perhaps not too surprising: The ``time''-trace of an
MC-sampled variable, at large values of step-size $l$, does exhibit a
sense of stiffness because a large proportion of moves is
rejected. Those moves that {\em are} accepted, will typically be large
and, at the same time, become particularly likely to be across the
bound state, see Fig.~2 of Ref.~\cite{10.2307/3182776}, thus creating
a sense of oscillation, though only in a statistical sense. We note
that if there is no bound state---as pertinent to uniform liquids, for
instance---Monte Carlo steps of {\em any} size, no matter how large,
are diffusive in nature, since $\lth = \infty$.  This notion can be
restated more formally using the standard Gaussian ansatz for the
density profile for a collection of particles,~\cite{dens_F1,
  doi:10.1080/00268978400101071} $\rho(\br) = \sum_i
(\alpha/\pi)^{3/2} e^{-\alpha (\br - \br_1)^2}$, where an individual
Gaussian $e^{-\alpha (\br - \br_1)^2}$ can be thought of as the
Boltzmann weight corresponding to a bounding potential centered at
$\br_i$. In the uniform liquid, $\alpha = 0$, the curvatures of the
bounding potentials vanish, hence $\lth \propto 1/\sqrt{\alpha} \to
\infty$.  Thus, automatically, simulations of uniform liquids cannot
be temporally calibrated against actual systems, nor could they be
regarded as computationally-efficient. Particle swaps, which can be
thought of as two simultaneous steps, do not approximate actual
motions, by construction. Thus MC simulations employing particle swaps
cannot be temporally calibrated against actual systems either, see
also below.

The properties of the transition matrix $\bpi$, which we examine next,
provide an instructive perspective on the preceding discussion of the
emergence of an optimal sampling rate. We have numerically
diagonalized the matrix corresponding to a discrete version of the
process described by Eq.~(\ref{MEcont}) for the parabolic energy
function (\ref{EHO}) in 1D.  As already mentioned, the spacing between
the largest and adjacent eigenvalues, $(\lambda_1 - \lambda_2) = (1 -
\lambda_2)$, gives the lowest relaxation rate or, equivalently, the
rate of sampling. The inverse of the latter spacing, i.e., the
corresponding relaxation time is shown in Fig.~\ref{llthFig} using
thin lines with dots. While the low- to moderate-$l$ portion of this
numerical result matches well the MC data, as well as the
analytically-derived asymptotics, the large-$l$ portion significantly
deviates from the correct value because of numerical issues. In this
parameter range, special care is needed to handle the smallness of the
Boltzmann weight $e^{-l^2/2 \lth^2}$, something we will not attempt
here.

\begin{figure}
  \includegraphics[width=0.8 \columnwidth]{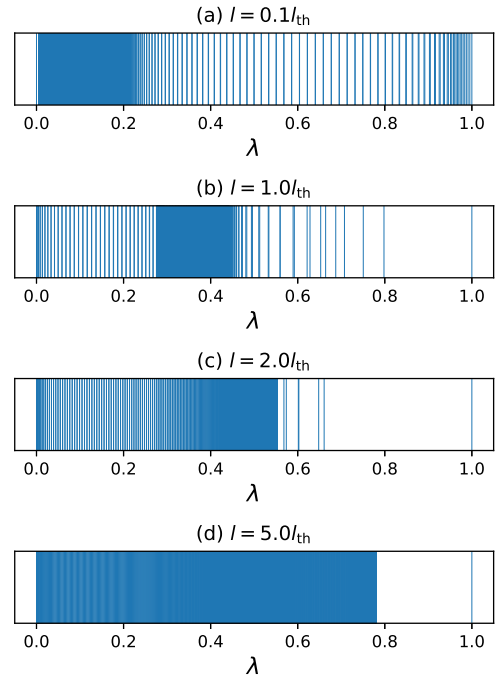}
  \caption{\label{gap} Graphical representation of the eigenvalue
    spectrum of the transition matrix $\bpi$ of the Markov chain for
    four different values of the $l/\lth$ ratio, for energy function
    (\ref{EHO}), $k=1$. Another representation, which makes the degree
    of degeneracy more clear, is provided in Fig.~\ref{gap2}. The
    matrix size is $1001 \times 1001$.}
\end{figure}

Already the existing data suffice to highlight the qualitative changes
in the eigen-value spectrum that accompany the diffusion-to-effusion
crossover: Direct inspection of the eigenvalues indicates that as one
approaches the crossover region coming from the diffusion limit $l \ll
\lth$, the spectra clearly develop a gap separating $\lambda_1$ and
$\lambda_2$, see Figs.~\ref{gap} and \ref{gap2}.  Note also that the
1D case analyzed above is the worst-case scenario in the sense that
the optimal rate has the biggest value for one-dimensional
simulations. Already in this worst-case scenario, the absolute value
of the apparent rate $(1-\lambda_2) \simeq 0.5$ is numerically close
to the value $-\ln \lambda_2$ it would have, if we used the discrete
prescription (\ref{psiMC}) for the wait times in place of the Poisson
statistics.  We provide values for both quantities $(1-\lambda_2)$ and
$-\ln \lambda_2$, respectively in Fig.~\ref{llthFig}, for the reader's
reference.

Conversely to the connections encapsulated in Eq.~(\ref{timemerged})
and (\ref{tomega}), we conclude that when multiple bound states with
distributed sizes are present, one cannot calibrate parameter $\Delta
t$ against an actual physical process. Despite this ambiguity, one
still can, in principle, calibrate {\em ratios} of the relaxation
rates for distinct bound states against actual systems, if it is known
that the attempted step length $l$ is shorter than the size $\lth$ for
every variety of bound state present.  Suppose for the sake of
argument that there are exactly two kinds of bound states, the
respective thermal lengths given by $\lth^{(1)}$ and
$\lth^{(2)}$. According to Eq.~(\ref{taullshort}), the ratio of the
corresponding sampling rates in the $l \to 0$ limit is given by
\begin{equation} \label{tau1tau2} \frac{\tau^{(1)}}{\tau^{(2)}} 
  = \left(\frac{\lth^{(2)}}{\lth^{(1)}} \right)^2, \hspace{3mm} l \ll
  \lth^{(1)}, \lth^{(2)}
\end{equation}
independent of the value of $l$ and $\Delta t$. The quantity on the
r.h.s. in the equation above is a geometric factor; it is also a ratio
of two equilibrium properties. We see that although this ratio is, in
principle, accessible to statistical sampling, the latter sampling
must be suboptimal. For a valid comparison with actual systems, such
systems must be overdamped.

\section{Classical Monte Carlo sequences
  are not physical trajectories, and ambiguities stemming from
  distribution of masses}
\label{contdiscr}

Despite a causal relationship among the sampled configurations,
sequences of coordinates generated in the course of Monte Carlo
simulation do not correspond to snapshots of individual physical
trajectories of inertial particles. Indeed, because of the absence of
inertia intrinsic to Gibbs-sampling of a Boltzmann distribution, one
is free to adopt a variety of probability distributions of waiting
times $t$, including those distributions that are non-vanishing as $t
\to 0$; the latter is the case for the Poisson statistics we employ
here. An elementary calculation shows that under these circumstances,
the distribution of the nominal rate of displacement $v_\tMC \equiv
x/t$ has a long, inverse-quadratic tail:
\begin{equation} p(v_\tMC) \propto \frac{1}{v_\tMC^2},
\end{equation}
and, hence, does not even have an average let alone higher moments.
Thus, Monte Carlo ``trajectories'' are inherently discontinuous
sequences of variables, while the displacement rate does not
correspond to a velocity of an inertial particle, notwithstanding the
kinematic-like relation (\ref{kinematic}) one may adopt under certain
circumstances.  Thus, even at optimal sampling---whereby statistical
sampling of relaxations might effectively mimic physical
motions---steps in classical Monte Carlo simulations can be, at best,
thought of as ordered in terms of a {\em pseudo}-time. We demonstrate
this more explicitly in what follows and, subsequently, begin
discussing implications for quantifying activated dynamics when
multiple bound states are present and masses of the bound modes are
distributed.

To drive home the distinction between the Monte Carlo step and a
physical time that can host inertial dynamics we consider a setup, in
which {\em both} the MC step and a true dynamical time are explicitly
present. First we recall that classical Monte Carlo simulations can be
thought of as a way to compute the configurational part of the
partition function of a system,
\begin{equation} Z = \int d x \:  e^{-\beta V(x)},
\end{equation}
specifically using importance sampling.~\cite{newman1999monte} Here,
$V(x)$ is the potential energy for the variable $x$.  Next we
consider, for the sake of argument, the configurational part of the
partition function for an infinitely large set of equivalent replicas
of a classical degree of freedom $x$ that is subject to a potential
energy $V(x)$:
\begin{equation} \label{Zr} Z_\text{r} =  \int dx \: Z(x, \beta)
\end{equation}
where
\begin{align}
  Z(x, \beta) & \equiv \lim_{N \to \infty} \int \left(
  \prod_{i=1}^{N-1} d x_i \right) \nonumber \\ & \times \exp\left\{ -
  \sum_{i=1}^N \delta t \left[ \frac{\kappa}{2}
    \frac{(x_{i}-x_{i-1})^2}{(\delta t)^2} + V(x_i) \right]
  \right\}  \nonumber \\ & = \int {\cal D}x \, \exp\left\{
  -\int_{-\beta/2}^{\beta/2} d t \left[ \frac{\kappa \dot{x}^2 }{2} +
    V(x) \right] \right\} \label{pathint}
\end{align}
and $\delta t \equiv \beta/N$. Each replica is coupled to exactly two
other replicas, thus allowing one to label these degrees of freedom
using ordinals $i=1, \ldots, N$ while imposing periodic boundary
conditions:
\begin{equation} \label{endpoints}
  x_{0} = x_{N} \equiv x.
\end{equation}
Thus the variables $x_i$, $i=1, \ldots, N$, are all formally
equivalent, per Eq.~(\ref{Zr}).  The replica-replica coupling
constant, $\kappa/2(\delta t)^2$, is constructed so that the integral
tends to a steady value in the limit $N \to \infty$. Indeed,
expression (\ref{pathint}) is formally equivalent to the diagonal
entries of the  (imaginary-time) density matrix for a {\em
  quantum} degree freedom $x$ of mass $m$ subject to a potential
$V(x)$~\cite{schulman2012techniques} where
\begin{equation} \label{kappam} \kappa \leftrightarrow \frac{m}{\hbar^2}.
\end{equation}

In any event, we use the importance sampling of the integrand in
Eq.~(\ref{pathint}) as a formal device to generate, in principle,
continuous trajectories. Indeed, for each given value of its
end-point, the sequence $ \{ x_i \} \leftrightarrow \{ x(t) \}$ is a
continuous function of the integration variable $t$ in the $N \to
\infty$ limit.  At the same time, an MC sequence for the end points
(\ref{endpoints}) of a individual trajectory can have arbitrary
increments, subject to the adopted proposal density. Next we stipulate
that the Gibbs-sampling of the distribution defined by the integrand
in Eq.~(\ref{pathint}) be optimal. This is equivalent to stipulating
that we take the path integral along the steepest descent, so as to
minimize the number of points where the integrand must be sampled.
Further, the trajectory $x_m(t)$ that maximizes the integrand
  in Eq.~(\ref{pathint}):
\begin{equation} \kappa \ddot x_m = -
  \prtl (-V)/\prtl x
\end{equation}
happens to be the Newtonian trajectory of a particle with inertial
mass $\kappa$ in the inverted potential $-V(x)$. Thus, the
integration variable $t$ in Eq.~(\ref{pathint}) is a true dynamical
time. The immediate vicinity of the optimum trajectory in the
direction of steepest descent determines a multiplicative factor for
the overall expression.

\begin{figure}
  \includegraphics[width=0.8 \columnwidth]{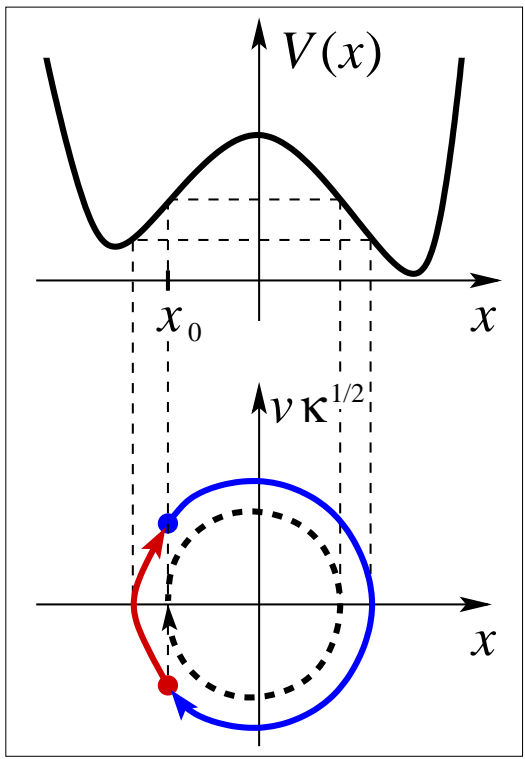}
  \caption{\label{twowell} Sketches of two types of stationary
    trajectories for density matrix (\ref{pathint}) for a bi-stable
    potential energy function: one trajectory is toward the adjacent
    bound state (thick red line), the other across the barrier
    separating distinct bound states (thick blue line).  The thick
    dashed line shows the limiting shape of the latter curve when
    $\kappa \to \infty$.}
\end{figure}

The $\kappa \to \infty$ limit in Eq.~(\ref{pathint}) corresponds to
the partition function sampled by {\em classical} Monte Carlo, by
virtue of the formal correspondence (\ref{kappam}).  The result of
taking this limit very much depends on the shape of the potential
energy $V(x)$. Consider, for the sake of concreteness, a two-well
potential and two distinct Newtonian trajectories---for the
  inverted potential $-V(x)$---each emanating from the same point
$x_0$, but having opposite signs for the velocity $v \equiv dx/dt$, as
in Fig.~\ref{twowell}.  We choose the point $x_0$ so that the
curvature $\prtl^2 V/\prtl x^2$ is positive at $x=x_0$, for
concreteness. It will be convenient to visualize the trajectories in
the plane $(x, v \sqrt{\kappa})$ (not the usual phase space $(x,
\kappa v)$!). The initial points for the two trajectories are shown
with dots. The two respective paths each have open ends, but smoothly
complement each other to a cyclic path.

As one takes the $\kappa \to \infty$ limit, the Newtonian trajectory
facing the adjacent bound state shrinks into a point at rates
proportional to $1/\kappa$ and $1/\sqrt{\kappa}$ along the horizontal
and vertical direction, respectively.  Also, the magnitude of
fluctuations around the classical path, along the $x$ direction,
diminishes at a rate $\propto 1/\sqrt{\kappa}$ for large values of
$\kappa$, in the usual fashion. Thus the path integral reduces to a
sampling of just one variable, i.e, the position of the endpoint
(\ref{endpoints}) itself when $\kappa \to \infty$, as is appropriate
in the classical limit.  The corresponding probability, up to the
first non-vanishing term in $1/\kappa$ is given by the expression:
\begin{equation} \label{Zbound} Z(x, \beta) \propto e^{-\beta [V(x) + 
      (\beta \prtl V/\prtl x)^2/12 \kappa]}
\end{equation}
where the prefactor depends on the detailed form of the potential. For
instance, for a parabolic bound state with a spring constant $k$, the
prefactor is given by the normalization factor of the classical
Boltzmann distribution $(\beta k/2\pi)^{1/2}$. Expression
(\ref{Zbound}) clearly tends to a finite value as $\kappa \to \infty$.

In contrast, the trajectory crossing the transition state, which
separates the two minima of $V(x)$, tends to a steady, cyclic shape in
the $\kappa \to \infty$ limit. This limiting shape, shown by the thick
dashed line in Fig.~\ref{twowell}, corresponds to the cyclic Newtonian
trajectory within the minimum of the inverted potential $-V(x)$ at
energy $E=-V(x_0)$ and represents an instanton.~\cite{Coleman1979}
Thus in contrast with the trajectory facing the adjacent bound state,
the path integral crossing the barrier corresponds to paths of finite
length even in the $\kappa \to \infty$ limit. The corresponding value
of the probability vanishes exponentially fast with $\kappa$ according
to the following, WKB-like expression:
\begin{equation}  \label{Zinstanton} Z(x, \beta) \propto e^{-\beta V(x)}
  e^{- 2 \int dx \sqrt{2 \kappa [V(x)-V(x_0)]}}.
\end{equation}
The vanishing of this expression in the $\kappa \to \infty$ limit
means that the assumption of the ability of optimal Monte Carlo
sampling to produce a continuous trajectory that connects two bound
states and, thus, is guaranteed to sample the corresponding transition
state is internally inconsistent. Conversely, optimal trajectories
must {\em avoid} transition states. Furthermore, the optimal step
length must be comparable to the distance between any two bound states
implying the sampling will be effusive and not exhibit activation
altogether.  Instead, the sampling rate will reflect the volume of the
thermally-populated region, as was argued in Section~\ref{FP}.

When suboptimal, Monte Carlo sampling can be used to explore both
bound states and transition states, of course, by making the typical
step size small enough and, thus preventing bypassing of
barriers. Still, in the absence of knowledge of the pertinent mass,
there is no guarantee that the so detected bound states are actually
present. Indeed, classical bound states in dimensions three and higher
are, generally, {\em not} robust against quantum
fluctuations.~\cite{schiff1968quantum} In spatial dimensions one and
two, attractive potentials exhibit at least one bound state, see
Ref.~\cite{doi:10.1063/1.1532538} and references therein. Still, the
wavelength of the corresponding motion is determined by the Broglie
wavelength corresponding to the depth $\epsilon$ of the bounding
potential, i.e. $\lambda \simeq \hbar/\sqrt{m \epsilon}$, a quantity
generally decoupled from the spatial extent of the bound state. (Note
one does not expect to find modes with a strictly vanishing mass in
dimensions one and two.~\cite{PhysRevLett.17.1133, PhysRev.158.383,
  Coleman1973}) In any event, one must separately decide whether the
effective mass of a mode exhibiting vibrational relaxation during MC
simulations is not so low as to lead an escape from the bound state by
tunneling. For example, it seems plausible that classical MC
simulations of water models might exhibit bound states that would be
actually escaped by the light-weight proton.

When tunnening is significant, the actual motions will be
qualitatively different from bound motions apparent to a classical
simulation.  For instance, imagine a two-well potential, each well
representing a classical bound state. As tunneling becomes
significant, with lowering the mass, a Larmor precession between the
two minima becomes possible, if the interaction with the environment
is not too strong.~\cite{PGW_QTST, doi:10.1063/1.449017, SB_review}
The frequency of the tunneling motion, then, sets a distinct,
intrinsically quantum time scale.  There is also the distinct
possibility that the bound states would all melt cooperatively
throughout the system, as would be the case during quantum melting of
a solid.~\cite{SchmalianWolynes2000, PhysRevB.68.134203, GHL, LKgap}
The threshold amount of tunneling can be thought of as separating just
two distinct phase behaviors that correspond to essentially classical
and quantum behaviors, respectively. (We note that the survival, if
any, of a bound state in the $m \to 0$ limit is analogous to the
survival of a replica-symmetry broken state~\cite{SpinGlassBeyond,
  2014arXiv1411.3941R} even as the replica-replica coupling vanishes.)
The formation of bound states can be viewed as a pseudo-transition $(m
=0) \to (m > 0)$; the transition can be either continuous or
discontinuous, by analogy with the metal-insulator
transition.~\cite{GHL, LKgap} Bound states represent instances of
broken translational symmetry. Thus one might view bound states and
the pseudo-time $t/\sqrt{m}$ as emerging together as a result of
lowering symmetry. In view of the slowness of the square root, one may
view the robustness---or lack thereof---of Monte Carlo-produced
relaxation functions, against the variation of mass among the modes,
as a question of how far from the $(m =0) \to (m > 0)$ transition the
system is, if the transition is continuous. (Trouton's law is an
important example of a similar robustness, the latter stemming from
the slowness of the logarithm.~\cite{LSurvey}) When the transition
happens to be discontinuous in the first place, one may view the
ambiguity with respect to mass variation, among the modes, as
quantitative, not qualitative. The more substantial source of
ambiguity has to do with the presence of multiple length scales in
problems of practical interest, as already stated, to be explicitly
illustrated next.

%\newpage

\section{Breaking of the pseudo-clock}
\label{examples}

Here we provide further evidence, using concrete model systems, that
the apparent connection between the pseudo-time emerging in an MC
simulation of a single-well bound state and physical timescales can
{\em not} be exploited, in practice, to quantify the kinetics of
activated transitions among distinct bound states. Such activated
transitions often underlie a variety of chemical and/or transport
phenomena of interest in applications. We shall call such activated
processes ``configurational'' to distinguish them from equilibration
within an individual free energy minimum and, also, in reference to
distinct free energy minima usually corresponding to distinct
microscopic configurations. Our first model system is a simple
bi-stable potential energy surface
\begin{equation} \label{bist} V(x) =  \frac{k}{2} [(x/x_0)^2 -1]^2.
\end{equation}
We fix the units by setting $x_0=1$ and $k=1$. The activation energy
$E^\ddagger = k/2$ for configurational equilibration due to inter-well
transitions is, thus, numerically equal to $1/2$. At sufficiently low
temperatures, one expects the rate of configurational relaxation to
become sufficiently lower than the rate of vibrational relaxation,
thus giving rise to a time-scale separation between the two dynamical
processes.  To elucidate a potential connection, if any, of these
actual dynamical processes with statistical sampling of the Boltzmann
distribution corresponding to the potential energy (\ref{bist}), we
perform Monte Carlo simulations for the latter potential energy using
the move
\begin{equation} \label{xmove}
  x \to y \hspace{10mm} \text{(``move type 1'')}
\end{equation}
where the value of the increment $(y-x)$ is distributed according to
the proposal density $q(x, y)$ from Eq.~(\ref{propdG}), as before. We
quantify the relaxation of the system by fitting the pair-correlation
function
\begin{equation} \label{xx}
  C_{xx}(t) \equiv \la x(t) x(0) \ra - \la x \ra^2 
\end{equation}
using a sum of exactly two exponential functions:
\begin{equation} \label{exp2} C_{xx}(t) = A_v e^{-t/\tau_v}
  + A_c e^{-t/\tau_c}, \hspace{2mm} \tau_v < \tau_c
\end{equation}
The labels ``$v$'' and ``$c$'' for the fitting parameters $\tau_v$ and
$\tau_c$ anticipate that the ``faster'' process has to do with
sampling an individual vibrational minimum while the ``slower''
process corresponds to sampling moves that traverse the maximum of the
potential energy $V(x)$ or, equivalently, the minimum of the bi-modal
Boltzmann distribution $p(x) \propto e^{-\beta V(x)}$.

\begin{figure}
  \includegraphics[width= 0.85 \columnwidth]{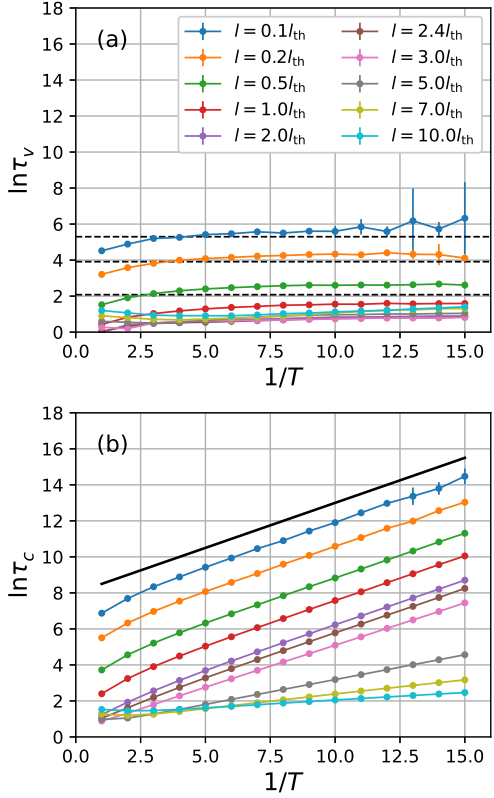}
  \caption{\label{dwelltaus} Arrhenius plots for the (a) shorter
    MC-relaxation time $\tau_v$ and (b) the longer MC-relaxation time
    $\tau_c$ for select values of the $l/\lth$ ratio obtained from
    two-exponential fits, Eq.~(\ref{exp2}), of the correlation
    functions (\ref{xx}) for energy function (\ref{bist}). $k=1$,
    $x_0=1$. In panel (a), the horizontal dashed lines show the value
    of the relaxation time evaluated using the short-$l$ asymptotics
    from Eq.~(\ref{taullshort}) for $l/\lth=0.1$, $0.2$, $0.5$ top to
    bottom, respectively. In panel (b), the straight line corresponds
    to an activation dependence for a barrier strictly equal to
    $k/2$.}
\end{figure}

\begin{figure}
  \includegraphics[width= 0.9
    \columnwidth]{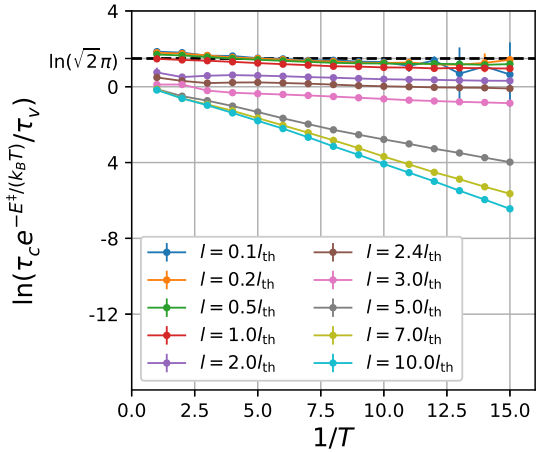}
  \caption{\label{dwelltauratio} Arrhenius plots for the quantity
    $\tau_c e^{-E^\ddagger/k_B T}/\tau_v$ for the relaxation times from
    Fig.~\ref{dwelltaus}.}
\end{figure}

The Arrhenius plots for the quantities $\tau_v$ and $\tau_c$ are shown
in Fig.~\ref{dwelltaus}(a) and (b), respectively.  The configurational
times are seen to follow rather closely the Arrhenius dependence for
sufficiently low values of the step $l$, but not so for step sizes
significantly greater than the size of an individual well, where the
sampling is in the effusive regime, consistent with the discussion
following Eq.~(\ref{Zinstanton}).  The prefactors for configurational
relaxation times clearly correlate with the corresponding vibrational
times. To quantify this correlation, we show the Arrhenius plot for
the quantity $\tau_c e^{-E^\ddagger/k_B T}/\tau_v$ in
Fig.~\ref{dwelltauratio}. In the first place, the dimensionless ratio
$\tau_c/\tau_v$ is worth plotting because it is independent of the
``time'' unit $\Delta t$ from Eq.~(\ref{psiMC}), thus making it,
essentially, a geometric quantity.

The data for $l/\lth \lesssim 1$ are seen to cluster around an
activated dependence corresponding to the expected activation barrier
$E^\ddagger = 1/2$ and, furthermore, appear to tend to a fixed value
at low temperatures. This behavior can be connected to some features
of activated dynamics for actual molecular systems in the presence of
strong damping. Indeed, we recall the simple approximation for the
activation rate for escape from a bound state in the overdamped,
Kramers limit discussed by Frauenfelder and Wolynes:~\cite{FW}
\begin{equation} \label{kesc} k_\text{escape} =
  \left( \frac{2 l_\smfp}{l_\sTS} \right) \frac{\omega}{2 \pi}
  e^{-E^\ddagger/k_B T}.
\end{equation}
The quantity $\omega$ is the vibrational frequency of the bound state,
while the $l_\smfp/l_\sTS$ ratio on the r.h.s. reflects a reduction of
the rate predicted by the transition state theory stemming from the
mean free path $l_\smfp$ being much smaller that the transition-state
size $l_\sTS$.~\cite{FW} The transition-state size is defined as the
size of the region containing all points that are within $k_B T$ worth
of energy from the barrier top.~\cite{FW} The relaxation rate in a
two-state system is equal to the sum of the two escape rates:
$1/\tau_c=k_{1 \to 2}+k_{2 \to 1}$, which amounts to twice the escape
rate from Eq.~(\ref{kesc}) in the present context. Thus one may use
Eqs.~(\ref{lthdef}) and (\ref{kinematic}) to obtain the following,
very simple expression
\begin{equation} \label{tvtc} \frac{\tau_c^\text{(ph)}
    e^{-E^\ddagger/k_B T}}{\tau_v^\text{(ph)}} = \frac{\pi}{2
    \sqrt{2}} \: \frac{l_\sTS}{\lth},
\end{equation}
where we used the label ``(ph)'' to emphasize that the above equation
pertains to a physical process. We see that in the overdamped limit,
the quantity in Eq.~(\ref{tvtc}) is not only geometric, as anticipated
above, but is also expressed exclusively in terms of equilibrium
quantities and, thus, might be accessible to thermodynamic sampling,
c.f. the discussion following Eq.~(\ref{tau1tau2}).  In more detail,
expression (\ref{tvtc}) is implicitly contingent on satisfying the
constraints $l_\smfp \ll l_\sTS$ and $\tau_\sauto \ll
\tau_v^\text{(ph)} \ll \tau_c^\text{(ph)}$, but it does not explicitly
contain either the mean free path $l_\smfp$ or the autocorrelation
time $\tau_\sauto$. At the same time, the Monte Carlo step size $l$
can be chosen to be smaller than the transition state size $l_\sTS$,
while the wait time $\Delta t$ of the semi-Markov process, from
Eq.~(\ref{psiMC}), explicitly cancels out in the $\tau_c/\tau_v$
ratio, as already mentioned. Thus, whenever the MC times obey $\tau_v
\ll \tau_c$ and the condition $l < l_\sTS$ is met, one should expect
the geometric relation (\ref{tvtc}) to also hold for the relaxation
profiles obtained in MC simulations, perhaps up to a factor of order
1, even though the MC times $\tau_v$ and $\tau_c$ cannot be
individually connected to actual microscopic times. This is because
expression (\ref{tvtc}) applies only when the actual dynamics is
overdamped, see the discussion following Eq.~(\ref{tomega}).

%Note the different scalings of the geometric factors with the thermal
%length in Eqs.~(\ref{tau1tau2}) and (\ref{tvtc}), respectively. The
%linear scaling in the latter equation comes about because of the
%characteristic linear scaling of the physical diffusivity
%(\ref{Dphys}) with the mean free path $l_\smfp$, which, in turn,
%stems from inertia, per Eq.~(\ref{kinematic}).

Now, the barrier top of the energy (\ref{bist}) is an inverted
parabola whose curvature $m (\omega^*)^2$ is twice less than the
curvature of the minima, hence $\omega^* = \omega/\sqrt{2}$.  Since
$l_\sTS = 2^{3/2} (k_B T/m)^{1/2}$,~\cite{FW}, we obtain that $l \ll
l_\sTS$ whenever $l \ll \lth$, and so the r.h.s. of Eq.~(\ref{tvtc})
should tend to $\pi \sqrt{2}$ at low temperatures.  This inference is
quantitatively consistent with Fig.~\ref{dwelltauratio}.

\begin{figure}
  \includegraphics[width= 0.9
    \columnwidth]{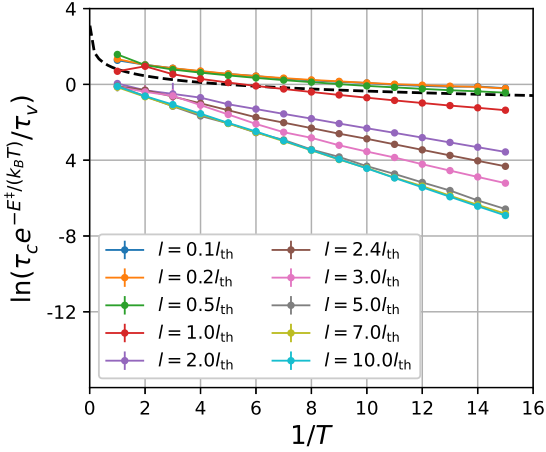}
  \caption{\label{cusptauratio} Arrhenius plots for the quantity
    $\tau_c e^{-E^\ddagger/k_B T}/\tau_v$ for the relaxation times for
    energy function (\ref{Ex}). $k=1$, $x_0=1$. The dashed line gives
    the numerical value of the r.h.s. of Eq.~(\ref{tvtc}).}
\end{figure}

Conversely, we can directly verify that when the condition $l <
l_\sTS$ is not satisfied, results of Monte Carlo simulation do not
strictly conform to Eq.~(\ref{tvtc}). Indeed, consider for the sake of
argument the following bi-stable energy function:
\begin{equation} \label{Ex} V(x) =  \frac{k}{2} (x^2 - x_0 |x| ),
\end{equation}
whereby the two minima are strictly parabolic while the barrier is a
cusp formed by two straight lines with slopes $\pm k x_0$. The
resulting data for the $\tau_c e^{-E^\ddagger/k_B T}/\tau_v$ ratio are
shown in Fig.~\ref{cusptauratio}. For this form of the potential, the
transition state size scales linearly with temperature $l_\sTS = 2
(k_B T/k x_0)$,~\cite{FW} and can become arbitrarily shorter than the
thermal length $\lth = \sqrt{k_B T/k}$. In any event, the Kramers
limit for actual dynamics would now yield $\pi \sqrt{k_B T/k}/x_0$ for
the r.h.s. of Eq.~(\ref{tvtc}).  Though this estimate falls in the
ballpark for the Monte Carlo-generated data in the low $l/\lth$
regime, we see that the simulation consistently underestimates the
activation barrier. This is expected because when $l > l_\sTS$,
crossing of the barrier is no longer conditional on sampling the
barrier top thus effectively leading to the barrier being bypassed,
the extent of bypassing depending on the value of $l/l_\sTS$.

The main message conveyed by Figs.~\ref{dwelltauratio} and
\ref{cusptauratio} is this: There is no reliable way to associate a
pseudo-time with activated relaxations observed in the course of Monte
Carlo simulations, unless the $l/\lth$ ratio is small and the barrier
top is smooth. Even so, such a small step size corresponds to a
diffusive sampling regime and is computationally inefficient.

We next elaborate on the preceding discussion of barrier bypassing. To
this end, we use an artificial energy function where the rate of
barrier-bypassing moves and the degree of bypassing can be rigidly
controlled. Indeed, consider the following simple energy function:
\begin{equation} \label{Exs} V(x, \sigma) = - (k x_0) x \sigma + \frac{k x^2}{2}
\end{equation}
where $x$ is a continuous degree of freedom while the quantity
$\sigma$ is an Ising spin-like degree of freedom that is allowed to
have only two values: $\sigma = \pm 1$. It is easy to see that the
effective free energy for the variable $x$:
\begin{align} \label{Fx} F(x) & = -k_B T \ln \left[
\sum_{\sigma = \pm 1} e^{-\beta V(x, \sigma)} \right] \nonumber \\ & =
  - k_B T \ln \left[ 2 \cosh (\beta k x_0 x) \right] + \frac{k x^2}{2}
\end{align}
becomes bistable below the temperature $T_0 = k x_0^2/k_B$, the two
minima corresponding to the vibrational ground states $x = \pm x_0$,
which in turn pertain to the two distinct values of the spin $\sigma =
\pm 1$, respectively. Thus one is able to unambiguously distinguish
the two alternative vibrational ground states of the system, despite
the barrier separating the two minima on the {\em free} energy surface
(\ref{Fx}) being finite. Note the $T \to 0$ limit of the free energy
$F(x)$ yields energy function (\ref{Ex}), up to an additive constant.
Energy function (\ref{Exs}) can be viewed as a classical limit of the
problem of non-adiabatic transport of a particle interacting with a
harmonic environment,~\cite{10.1143/PTP.13.160, OvchOvch,
  doi:10.1146/annurev.pc.15.100164.001103} whereby the two terms
corresponding to $\sigma=\pm 1$, respectively, are the usual Marcus
parabolas.

\begin{figure}
  \includegraphics[width=0.7 \columnwidth]{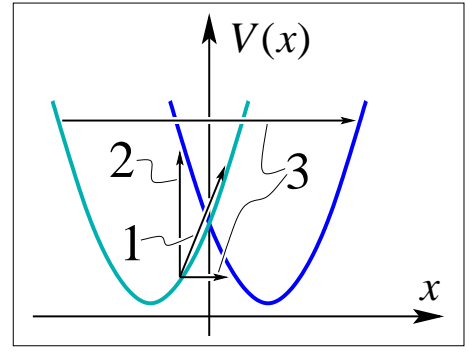}
  \caption{\label{moves} Monte Carlo moves of the three types employed
    in the present work, Eqs.~(\ref{xmove}), (\ref{sflip}), and
    (\ref{pseudot}) and the respective energy changes are graphically
    illustrated by straight arrows.}
\end{figure}

\begin{figure*} \centering
  \includegraphics[width= 1.7 \columnwidth]{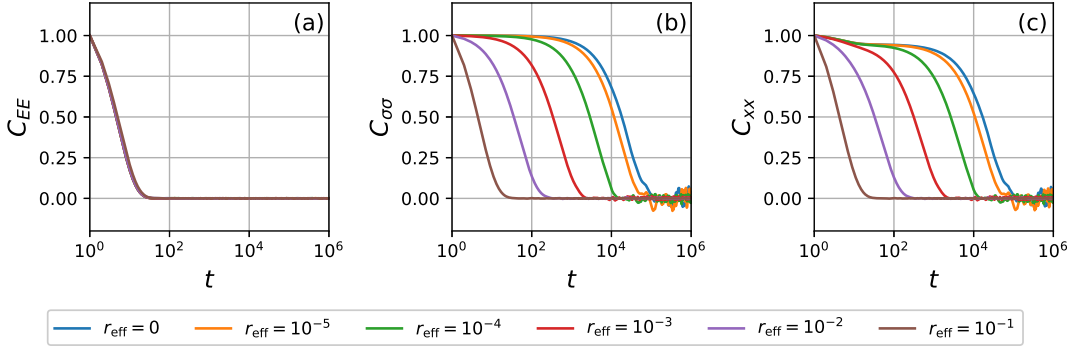}
  \caption{\label{activated1} Relaxation profiles for equilibrium
    Monte-Carlo simulations of energy function (\ref{Exs}). $T =
    0.07$, $k = 1$, $l/\lth = 2.38$, $r=1$. Panels (a), (b), and (c),
    show the correlation functions from Eqs.~(\ref{EE}), (\ref{ss}),
    and (\ref{xx}), respectively.}
\end{figure*}

We will continue using moves of type 1 from Eq.~(\ref{xmove}) to
sample the individual vibrational terms. To enable transitions between
the latter terms, we will use a combination of two types of move: One
is the simple spin-flip move:
\begin{equation} \label{sflip} \sigma \to - \sigma \hspace{10mm}
  \text{(``move type 2'')}
\end{equation}
Already moves of type 1 and 2, together, allow one to fully sample the
phase space of the system. By construction, we denote the number of
attempts for move 2, relative to move 1, with letter $r$. We directly
illustrate in Fig.~\ref{moves} that the energy cost to transition
between the two vibrational ground states is bounded from below by the
barrier height $E^\ddagger = k/2$, if only moves of type 1 and 2 are
used.

The other move type that will be used to switch between the
vibrational terms involves {\em both} variables at the same time:
\begin{equation} \label{pseudot}  x \to -x, \: \: \sigma \to -
\sigma \hspace{10mm} \text{(``move type 3'')}
\end{equation}
This move is isoenergetic and, thus, bypasses the classical barrier
separating the two minima of the energy function, if the initial
energy is below the crossing point of the two Marcus parabolas, see
Fig.~\ref{moves}. A combination of moves of type 1 and 3 is fully
ergodic. We will denote the rate of these ``non-physical'' moves,
relative to the rate of moves of type 1, with $r_\text{eff}$, in
reference to these moves being strictly effusive.

In Fig.~\ref{activated1}, we display the following pairwise
correlation functions:
\begin{align} \label{EE}
  C_{EE}(t) & \equiv \la E(t) E(0) \ra - \la E \ra^2 \\
  \label{ss} C_{\sigma \sigma}(t) & \equiv \la \sigma(t) \sigma(0) \ra
  - \la \sigma \ra^2
\end{align}
as well as the correlation function (\ref{xx}). Data in
Fig.~\ref{activated1} cover a broad range of the attempt rate
$r_\text{eff}$ for the non-physical move.

We observe that the energy-energy correlation function relaxes much
faster than the other two correlation functions. This is because the
two vibrational ground states are degenerate and so the value of
energy is not sensitive to the precise identity of the minimum. In
contrast, the spin-spin correlation function does not reach its
long-term value until the configurational equilibration has
occurred. Already above a certain, very low rate of the non-physical
move of type 3, the kinetics of configurational relaxation are
dominated by the latter non-physical process. Furthermore, above a
certain value of $r_\text{eff}$, the overall relaxation cannot be
identified with either vibrational or activated processes and is
completely slaved to the effusive move. We also observe that in
addition to short-circuiting the activation barrier, the non-physical
move modifies the vibrational relaxation, too.

From here on, we focus on the activated dynamics and set
$r_\text{eff}=0$ until further notice.  Both the vibrational and
configurational contributions are clearly seen in the $x-x$
correlation function, whereby the configurational contribution shows
up as a pronounced plateau.  The vertical position of the plateau is a
Debye-Waller factor. Lowering of this position, relative to unity, is
approximately given by the $\lth/x_0$ ratio, of course.  Note that the
spin-relaxation profile {\em also} has a shorter-term, vibrational
contribution, but the latter contribution is determined by the
staggered susceptibility of the spin, $\simeq 1/\cosh^2(\beta k
x_0^2)$, and is too small numerically to be seen on the graph, at the
temperature in question.

\begin{figure}
  \includegraphics[width= \columnwidth]{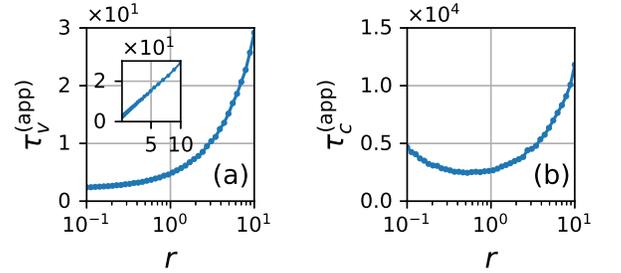}
  \caption{\label{taubare} Apparent relaxation times determined by
    fitting the time traces of the $C_{xx}$ from Eq.~(\ref{xx}), using
    the two-exponential form from Eq.~(\ref{exp2}), as functions of
    the ratio $r$ of the attempt rates for the configurational and
    vibrational degree of freedom. Panels (a) and (b) show the shorter
    time $\tau_v$ and longer time $\tau_c$, respectively. The energy
    function is from Eq.~(\ref{Exs}). $T = 0.1$, $k = 1$, $l/\lth =
    2.38$. }
\end{figure}

\begin{figure}
  \includegraphics[width= \columnwidth]{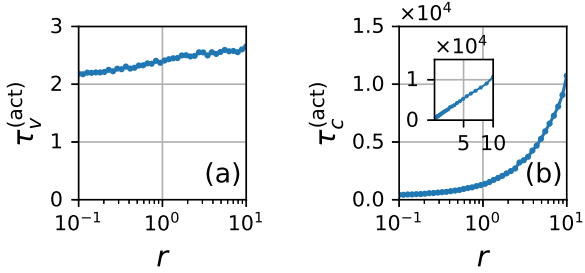}
  \caption{\label{taur} Relaxation times from Fig.~\ref{taubare}
    rescaled according to Eqs.~(\ref{r1}) (panel a) and (\ref{r2})
    (panel b), respectively, so as to that account for the variation
    of the attempt rate $r$.}
\end{figure}

In Fig.~\ref{taubare}, we show the $r$ dependence of the apparent
relaxation times $\tau^\text{(app)}_v$ and $\tau^\text{(app)}_c$ for
the vibrational and configurational relaxation, respectively, as
inferred directly from the relaxation profiles.  The near linear
increase of $\tau^\text{(app)}_v$ with the attempt ratio $r$, seen in
Fig.~\ref{taubare}, occurs because for greater values of $r$, a
progressively larger fraction of attempts is spent on the spin flip,
and vice versa for $\tau^\text{(app)}_c$.  Thus in Fig.~\ref{taur}, we
graph the relaxation times suitably adjusted to accommodate for this
circumstance:
\begin{align} \label{r1} \tau^\text{(act)}_v & = \tau^\text{(app)}_v \frac{1}{1+r} \\
  \tau^\text{(act)}_c & = \tau^\text{(app)}_c \frac{r}{1+r} \label{r2}
\end{align}
We observe that the rescaled vibrational times now vary only modestly
within a substantial range of $r$. At the same time the rescaled
configurational time varies about as much as the apparent time
$\tau^\text{(app)}_c$. In any event, both $\tau^\text{(act)}_v$ and
$\tau^\text{(act)}_c$ are seen to increase monotonically with $r$.  We
observe that even when the number of control parameters for the
simulation is the same as the number of distinct relaxation times, the
latter relaxation times cannot be optimized simultaneously. This
notion is consistent with the discussion following
Eq.~(\ref{Zinstanton}).

\begin{figure}
  \includegraphics[width= 0.9 \columnwidth]{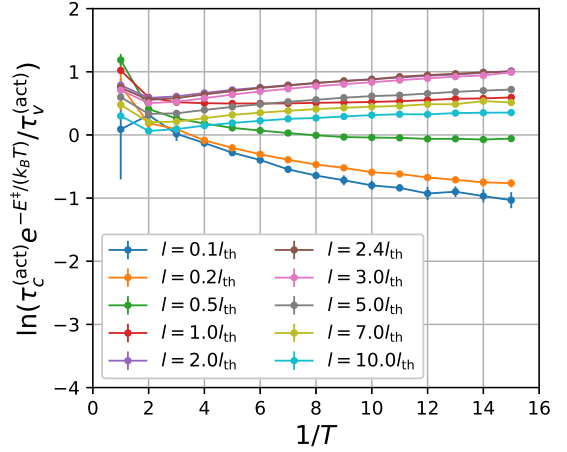}
  \caption{\label{dwelltauratiocusp} Arrhenius plots for the quantity
    $\tau_c e^{-E^\ddagger/k_B T}/\tau_v$ for the relaxation times
    observed during Gibbs-sampling energy function (\ref{Exs}).  The
    times are rescaled to account for the variation of the attempt
    rate $r$, as in Fig.~\ref{taur}. $k = 1$, $r=0.5$. A straight line
    with a vanishing slope would correspond to strict activation at
    $E^\ddagger = k/2 = 1.2$.}
\end{figure}

\begin{figure}
  \includegraphics[width= \columnwidth]{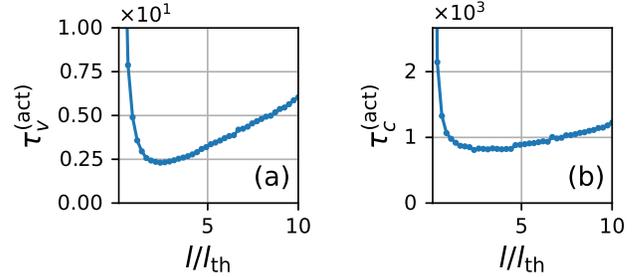}
  \caption{\label{taul} Rescaled relaxation times from Eqs.~(\ref{r1})
    (panel a) and (\ref{r2}) (panel b) as functions of the step length
    $l$.  $k = 0.1$, $r=0.5$.}
\end{figure}

Next we show in Fig.~\ref{dwelltauratiocusp} the simulated value for
the quantity $\tau_c e^{-E^\ddagger/k_B T}/\tau_v$ for the energy
function (\ref{Exs}). Here we observe that in contrast with the
bi-stable potential energy functions analyzed earlier, the quantity
$\tau_c e^{-E^\ddagger/k_B T}/\tau_v$ now depends on the $l/\lth$
ratio in a decidedly non-monotonic fashion. This behavior stems from
the recurrent dependence of the relaxation time within an individual
bound state. The configurational relaxation time also follows this
trend, see Fig.~\ref{taul}, but to a lesser degree. In any event, we
observe that for all values of the $l/\lth$ ratio, the data in
Fig.~\ref{dwelltauratiocusp} correspond to a lack of strict
activation, within the temperature range of the simulation. A strict
Arrhenius dependence does appear to set in at lower temperature, at
the expected value of the activation barrier $E^\ddagger = 1/2$,
consistent with the lack of barrier-bypassing when $r_\text{eff}=0$.

\begin{figure}
  \includegraphics[width= .85 \columnwidth]{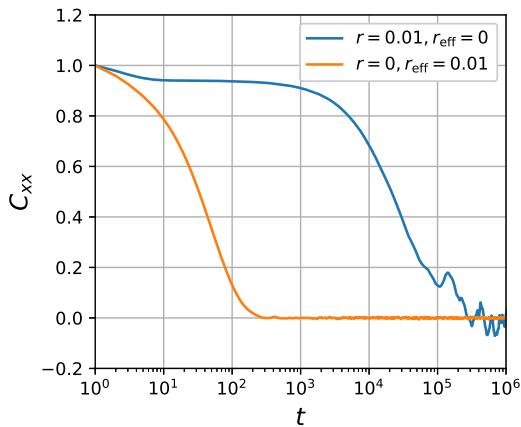}
  \caption{\label{protocols} Select relaxation profiles for
    equilibrium Monte-Carlo simulations of energy function (\ref{Exs})
    for the ``physical'' and ``non-physical'' protocols, respectively,
    as pertinent to Figs.~\ref{sampling1} and \ref{sampling2}. $T =
    0.1$, $k = 1$, $l/\lth = 2.38$.}
\end{figure}

In the remainder of this Section, we consider effects of using
barrier-bypassing moves on the quality of sampling of the free energy
surface. We focus on two distinct simulation protocols, both
ergodic. Protocol 1, call it the ``physical'' protocol, uses only
moves of type 1 and 2, whereby bypassing of the barrier is strictly
forbidden. Protocol 2, call it the ``non-physical'' protocol, uses
only moves of type 1 and 3. Equilibrium relaxation profiles, as well
as the pertinent parameters are provided in Fig.~\ref{protocols} and
its caption. We choose the attempt rate of move 3 so that the
corresponding relaxation is much faster than that in protocol 1, but
still takes a substantial number of Monte Carlo steps, i.e., 50 or so.

\begin{figure}
  \includegraphics[width= .9 \columnwidth]{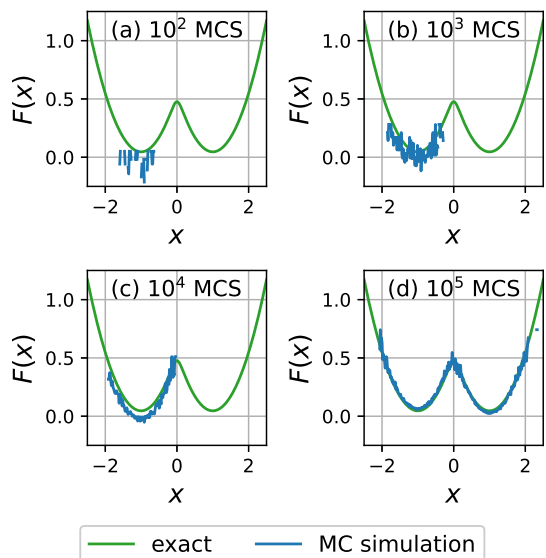}
  \caption{\label{sampling1} Histogram of the values of the coordinate
    $x$ for four select values of the collection time, for the
    ``physical'' protocol, move types 1 and 2, energy function from
    Eq.~(\ref{Exs}), same conditions as in Fig.~\ref{protocols}. The
    solid line is the equilibrium free energy as a function of $x$,
    from Eq.~(\ref{Fx}).}
\end{figure}

\begin{figure}
  \includegraphics[width= .9 \columnwidth]{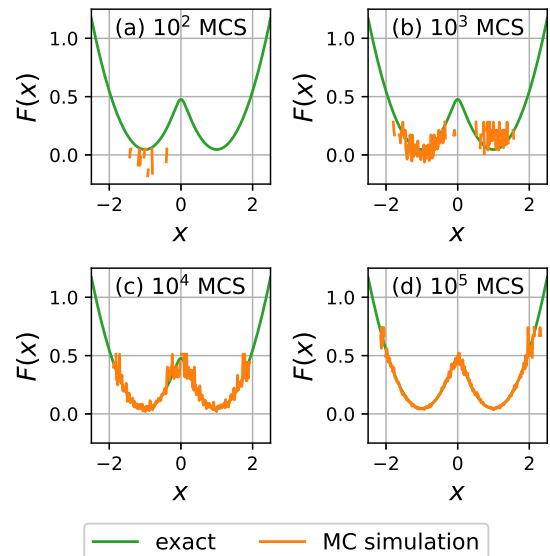}
  \caption{\label{sampling2} Histogram of the values of the coordinate
    $x$ for four select values of the collection time, for the
    ``non-physical'' protocol, move types 1 and 3, energy function
    from Eq.~(\ref{Exs}), same conditions as in
    Fig.~\ref{protocols}. The solid line is the equilibrium free
    energy as a function of $x$, from Eq.~(\ref{Fx}).}
\end{figure}

In Figs.~\ref{sampling1} and \ref{sampling2}, we histogram the values
of the coordinate $x$ that we collect starting in some equilibrated
configuration. We present results for four select values of the
collection time, where we normalize the histograms and compute the
corresponding free energy by taking the logarithm and multiplying by
$(-k_B T)$.  We observe that for the physical protocol, both free
energy minima become adequately sampled on times comparable to the
typical relaxation time. The transition-state region will have been
sampled by that point as well.  These results can be contrasted with
those obtained using the non-physical protocol, see
Fig.~\ref{sampling2}: When the collection times exceed the relaxation
time by an order of magnitude or so, both minima have been securely
sampled, but not the transition state. The latter only becomes
adequately sampled, at least by visual inspection, on times that are
at least two orders of magnitude greater than the relaxation time.

To rephrase these notions from the viewpoint of computational
efficiency, the non-physical protocol offers an improvement in the
efficiency of thermodynamic sampling by about two orders of magnitude,
but it is achieved at the expense of avoiding transition state
configurations, which are relatively high in free energy. While
employment of the non-physical move does seem to speed up the sampling
of the transition-state configurations, the improvement is only
modest, an order of magnitude or so. Most importantly, the timescale
on which the transition state becomes adequately sampled exceeds the
apparent relaxation time by at least two orders of magnitude.  We
expect these trends to be general, the quality of transition state
sampling becoming progressively poorer for higher barriers. Such
accelerated simulations are unlikely to generate the bottle-neck
configurations on which the transport is conditional, thus preventing
one from establishing the mechanism of the activated transitions.
This is, of course, consistent with the absence of slow processes, in
the simulated relaxation profiles, that one associates with activation
in the first place.

\section{Brief summary and implications for simulations of glassy mixtures}
\label{conclusions}

We have shown that notwithstanding its ability to efficiently sample
the thermodynamics of a classical system, Gibbs sampling of the
Boltzmann distribution cannot be used to reliably quantify the
kinetics of relaxation toward equilibrium. Because Gibbs sampling
lacks an inherent time scale, one cannot unambiguously assign a time
scale to the semi-Markov process associated with the sampling, if the
original physical system exhibits a distribution of relaxation rates.

The argument started out by considering the special case when the
physical motion is confined to a single-well bound state in
equilibrium with its environment. When not overly damped, this motion
can be mimicked by a semi-Markov process, if the sampling is
optimal. This setup is rather peculiar and seems to pertain to
mechanically stable arrangements of hard objects.  Hereby the
relaxation rate of MC simulation can be unambiguously connected with
the actual relaxation rate and frequency of the vibrational motion
within the bound state, via a universal relation.  Also at optimal
sampling, the transition matrix of the Markov chain corresponding to
the Metropolis sampling develops a well defined gap between the two
largest eigenvalues. One may thus define a pseudo-clock, which can be
thought of as emerging together with the underlying bound state and
thus requires symmetry breaking.  In this way the pseudo-clock is
similar to actual clocks based on cyclic processes, since such cyclic
motions {\em also} require symmetry breaking to occur so as to confine
the motion. Furthemore, when the clock is based on a quantum process,
the corresponding frequency is directly related to a gap in the
spectrum as is the pseudo relaxation time.

%For shorter-than-optimal step sizes, the Monte Carlo sampling becomes
%more diffusive in nature, while for longer-than-optimal paths, the
%sampling process is similar to {\em effusion}, where the typical
%states occupy a small patch is the space of configurations.

In all other cases, the pseudo-clock breaks down. For instance, MC
simulations of uniform liquids, which are translationally invariant,
cannot be calibrated against actual systems.  When two or more bound
states are present, a number of distinct relaxation processes will
take place. Since it is impossible to optimize the step size for all
processes at the same time, the dynamics for the compound process will
be rendered incorrectly. Of particular interest is the interplay of
activated relaxation among distinct bound states and the vibrational
relaxation within in an individual bound state.  We have shown
explicitly that vibrational relaxation---as apparent in MC
simulations---cannot be reliably used as a reference time scale for
activated reconfigurations among the bound states.  This is
particularly evident in the effusive regime of sampling, whereby the
step size can be made so large as to cover more than one bound
state. Hereby, transitions among distinct bound states do not require
activation because the corresponding barriers are bypassed.  Such
effusive moves restore the symmetry while the relaxation time is
determined by the volume of the thermally accessible regions, not the
height of the barrier separating the bound states.  Making the step
size small allows one to recover a time-scale separation between
vibrational and configurational equilibration, as well as the height
of the activation barrier, but at the cost of making the simulation
diffusive in nature and, thus, impractically slow. In any event, the
values of the rates of configurational and vibrational relaxation
alike can not be individually connected with actual physical
processes.

According to the present findings, accelerated Monte Carlo protocols
that make thermodynamic sampling more efficient do so at the expense
of avoiding the bottle-neck configurations corresponding to the
transition states for activated transport thus precluding one from
detecting the latter configurations.

%We have also observed that as bypassing barriers becomes more
%efficient, vibrational relaxations within individual minima becomes
%modified, too. We have constructed an explicit example where
%relaxation in a two-well bound state cannot be identified with a
%physical process altogether when sampling becomes dominated by
%effusive moves.

The latter notion has implications for simulational studies of glassy
liquids; some of these studies were mentioned in the Introduction.
According to the RFOT theory,~\cite{KTW, LW_ARPC, L_AP} molecular
transport in glassy liquids and frozen glasses is slow because the
translational symmetry characteristic of uniform liquids is
transiently broken on times shorter than the lifetimes of metastable
aperiodic structures that form below a dynamic crossover.  Transport
in such liquids requires activated escape from those metastable
configurations and, thus, is subject to bottle-neck, transition-state
configurations, in which a relatively large number of particles, call
it $N^\ddagger$, must reconfigure.~\cite{LW_aging, LRactivated} The
quantity $N^\ddagger$ grows as temperature is lowered, implying
activated transport in glassy liquids should become increasingly more
cooperative alongside. In quantitative terms, the cooperativity scale
$N^\ddagger$ grows to about 50 rigid molecular units near the
laboratory glass transition on the hour time scale.~\cite{XW,
  RWLbarrier, LRactivated} Once the the bottle-neck configuration is
overcome, the transition will proceed to span a relatively compact
region of size $N^* \simeq 4 N^\ddagger$.~\cite{XW, EastwoodW,
  Lrelics} Glassy liquids are predicted to exhibit inherent built-in
stress~\cite{BL_6Spin, BLelast} that varies spatially on the length
scale $(N^*)^{1/3}$; it is however not clear whether there should be
an explicit spatial signature to this variation of stress. Even so, it
was argued in Refs.~\cite{ZL_JCP, ZLMicro2, LL2} that in materials
exhibiting charge-density waves, the strained regions exhibit a
quantum-mechanical signature in the form of midgap electronic states.

According to the present results, accelerated Monte Carlo protocols
that avoid bottle-neck configurations for activated transport cannot
be used to quantify the extent of cooperativity during the activated
events. We have devised a caricature model that has common features
with particle-swap simulations. In both cases, two degrees of freedom
undergo large, effusive displacements at the same time, a non-physical
move that can be used to dramatically speed up thermodynamic sampling
of typical states. We have seen that the deeper in the free energy
landscape the accelerated simulation can equilibrate, the less likely
is one to detect transition-state configurations.  Considering that
effusion is a phenomenon pertinent to dilute gases, barrier-bypassing
moves---when used to simulate a glassy liquid---can be thought of as
the liquid boiling locally! Such processes are clearly physically
irrelevant at densities in question. We have also seen that such
effusive processes can modify properties of the bound states and can
even effectively destroy them.

Conversely, the present results indicate that to quantify the
cooperativity and kinetics of activated reconfigurations in a glassy
liquid, one must impose a sense of continuity to the trajectories at
least to some extent. This can be done by (numerically) solving
Newton's equations of motion, of course. Alternatively, one can employ
Monte Carlo simulations in the diffusive regime, as already mentioned,
or by simulating multiple, coupled replicas of the system, per the
discussion in Section~\ref{contdiscr}. In all of those cases, however,
the respective protocols are computationally expensive. The notion of
replicating the system---with the aim of elucidating activated
transport---should not be too surprising, since replica methodologies
can be used to detect breaking of translational symmetry in aperiodic
systems that do not have an obvious structural-reference
state.~\cite{PhysRevLett.75.2847, 2014arXiv1411.3941R}

Likewise, the present results are consistent with a notion that
hydrodynamic, coarse-grained descriptions of the glass transition,
such as the mode-mode coupling theory,~\cite{Goetze2008complex} (MCT)
do not apply below the dynamic crossover. Indeed, such coarse-graining
implies that system can bypass the reconfiguration barriers thus
leading to a description that effectively operates on a single free
energy minimum. In contrast, the number of the latter minima scales
exponentially with the system size and becomes huge already for
modestly-sized samples, a fact that can be directly confirmed using
calorimetry.~\cite{L_AP}

\subsection*{Acknowledgments}  We thank Eric Bittner for sharing
his expertise.  We gratefully acknowledge the support by the NSF
Grants CHE-1465125 and CHE-1956389, the Welch Foundation Grant E-1765,
and a grant from the Texas Center for Superconductivity at the
University of Houston. We gratefully acknowledge the use of the
Maxwell/Opuntia Cluster at the University of Houston.  Partial support
for this work was provided by resources of the uHPC cluster managed by
the University of Houston and acquired through NSF Award Number
ACI-1531814.  This article is dedicated to the memory of Hans
Frauenfelder.

\appendix

\section{General expression for the  relaxation profile of an eigenvector
  of the transition matrix for an arbitrary distribution of wait
  times}
\label{psitspec}

The following discussion heavily relies on standard results from
renewal theory,~\cite{Ross} however is intended to be reasonably
self-contained. Everywhere below, the quantity $t$ stands for the time
variable of semi-Markov processes, not a physical time. Suppose the
initial condition $\bp(t=0)$ for the Markov chain is given by a
non-stationary eigenvector of the transition matrix $\bpi$ whose
eigenvalue is $\lambda < 1$. Following each Monte Carlo event, the
transition matrix is applied to the current value $\bp(t)$. $n$ such
applications to the initial distribution results in scaling it down by
an overall factor $\lambda^n$. By time $t$, an arbitrary number of
Monte Carlo can occur, in principle, depending on the detailed form of
the waiting-time distribution $\psi(t)$. Here we ask: What is the
expectation value for the scaling factor $p_\lambda(t)$, at time $t$?
By construction:
\begin{equation} \bp(t) = p_\lambda(t) \bp(t=0).
\end{equation}
We determine this expectation value $p_\lambda(t)$ by averaging
$\lambda^n$ over the probability $P_n(t)$ that exactly $n$ Monte Carlo
events occurred by time $t$:
\begin{equation} \label{pl1}
  p_\lambda(t) = \sum_{n=0}^\infty \lambda^n P_n(t).
\end{equation}
Consider, for instance, the Poisson process with relaxation time
$\Delta t$: $P_n(t) = e^{-t/\Delta t}(t/\Delta t)^n/(n!)$. In the
latter case, Eq.~(\ref{pl1}) immediately yields $p_\lambda(t) =
e^{-(1-\lambda) t/\Delta t}$, consistent with
Eq.~(\ref{Ptuni1}). Generally,
\begin{equation} \label{pnt}
  P_n(t) = F_n(t)-F_{n+1}(t),
\end{equation}
where $F_n(t)$ is the (cumulative) probability that the $n$-th renewal
has occurred by time $t$ or, equivalently, that the number of renewals
by time $t$ is greater than or equal to $n$. By construction, the
survival probability of the renewal process is the probability that no
renewal has occurred:
\begin{equation}  \label{survp}
  P_0(t) = 1 - F_1(t) = 1 - \int_0^t \psi(t) \: dt
\end{equation}
and we avoid pathological cases by adopting
\begin{equation} \label{patho} F_n(0) = 0.
\end{equation}
The probabilities $F_n(t)$ are connected through an iterative
relation:
\begin{align} F_{n+1}(t) & = \int_{t_n + t_1 < t} d F_n(t_{n}) \: d F_1 (t_1)
\\ &  = \int_0^t F_n(t-t_1) \: d F_1 (t_1),
\end{align}
which, we see, happens to be a convolution. Taking the Laplace
transform and using the second equality in Eq.~(\ref{survp}), as well
as Eq.~(\ref{patho}), readily yields:
\begin{equation} \widetilde{F}_n(s) = \frac{1}{s} \, \tps^n(s).
\end{equation}
Computing the Laplace transform of Eq.~(\ref{pl1}), while using
Eq.~(\ref{pnt}), becomes a matter of summing a geometric series,
which, then, yields Eq.~(\ref{plambdat}) of the main text.

\section{Diffusion-to-effusion crossover: Auxiliary information}
\label{effusion}

First we obtain an approximation for the Monte Carlo sampling rate in
the large $l$ limit of Eq.~(\ref{contmod}) for the one-dimensional
case. We do not have in our possession the functional form of
eigenvectors pertaining to the largest non-stationary eigenvalue
$\lambda$ of the transition matrix. Instead, we consider a simple
trial form that is linearly independent from the stationary solution
of master equation (\ref{contmod}), $\tp=\text{const}$, and whose
symmetry with respect to reflection about the origin is consistent
with the energy function (\ref{EHO}).  Specifically we consider a
solution that is odd and has its single node located at the
origin---while being constant on the positive and negative side of the
vertical axis, respectively:
\begin{equation} \label{ansatz} p(x, t) = A(t) \, \text{sign}(x)
\end{equation}
With the aid of this ansatz, proposal density (\ref{propdG}) and
energy function (\ref{EHO}) yield for $x<0$:
\begin{equation} \label{incons} (\Delta t) \frac{- \dot{A}}{2 A}
  = \int_0^{|x|} dy \, q(x, y) + \int_{|x|}^\infty dy \,
  e^{-(y^2-x^2)/2 \lth^2} \: q(x, y)
\end{equation}
Since the r.h.s. of this equation varies with $x$, the ansatz from
Eq.~(\ref{ansatz}) is internally-inconsistent, the degree of
inconsistency dependent on the value of $x$. Still, Eq.~(\ref{incons})
can thought of, informally, as providing the relaxation rate as a
function of the coordinate. Averaging both sides of the equation with
respect to the equilibrium distribution of $x$ will single out the
most relevant values of the coordinate. Thus, we multiply by $2
e^{-x^2/2 \lth^2}/(2 \pi \lth^2)^{1/2}$ and integrate over the
negative $x$'s:
\begin{align} \label{rate} -\frac{\dot{A}}{A}
  = \frac{4}{\Delta t} \int_{-\infty}^0 \frac{dx}{\sqrt{2\pi \lth^2}}
  \left[e^{-x^2/2 \lth^2} \int_0^{|x|} dy  \: q(x, y)
    \right. \nonumber \\ + \left. \int_{|x|}^\infty dy \, e^{-y^2/2
      \lth^2} \: q(x, y) \right]
\end{align}
In the limit $\lth \ll l$, one can ignore the variation of the
function $q(x, y)$ in both integrals over $y$: $q(x, y) \approx
1/\sqrt{2\pi l^2}$. Thus, the first integral is well approximated by
$|x|/\sqrt{2\pi l^2}$. The second integral will be approximated by the
expression $e^{-x^2/2 \lth^2} (\lth/2 l )$, which is viewed as a
compromise between the small-$x$ power-law expansion and the
large-$x$, asymptotic expansion of the error function. (The resulting
error, if any, becomes less significant in higher dimensions, where
this term is subdominant, see below.) This immediately gives:
\begin{equation} -\frac{\dot{A}}{A} (\Delta t) \approx \left( \frac{2}{\pi} +
  1 \right) \frac{\lth}{l} \approx \frac{1}{0.61} \: \frac{\lth}{l}.
\end{equation}

Higher-dimensional cases can be considered analogously, by making the
function $p$ odd along exactly one spatial direction. This yields
\begin{equation} \label{taulllongd} -\frac{\dot{A}}{A} (\Delta t) \approx
  \left[ \frac{2}{d} \: \frac{\Gamma(d)}{\Gamma^2(d/2)} + 1 \right]
  \left( \frac{\lth}{l} \right)^d \equiv f_d \left( \frac{\lth}{l}
  \right)^d,
\end{equation}
where $\Gamma$ is the standard gamma function.~\cite{AS}

We next introduce a simple interpolative expression for the
$(l/\lth)$-dependence of the sampling rate that is consistent with
both the small and the large $l$ asymptotics, Eqs.~(\ref{taullshort})
and (\ref{taulllongd}) respectively, as well as the $d \to \infty$
scaling of the optimal value of $l$.
\begin{equation} \label{taull} \frac{\Delta t}{\tau} \simeq \frac{1}{2}
  \frac{(l/\lth)^2 }{[n_d (l/\lth)^2 + 1]^{d/2 + 1}}.
\end{equation}
where 
\begin{equation} n_d \equiv (2 f_d)^{-2/(d+2)}
\end{equation}
and $f_d$ is defined in Eq.~(\ref{taulllongd}).

Eq.~(\ref{taull}) readily yields for the optimal step size:
\begin{equation} \label{llopt} \frac{l_\text{opt}}{\lth} = \frac{1}{\sqrt{d}} \:
  \left[2^{1/2} (2 f_d)^{1/(d+2)} \right] \equiv
  \frac{C_1 (d)}{d^{1/2}} \xrightarrow[d \to \infty]{}
  \frac{2^{3/2}}{\sqrt{d}}
\end{equation}
The quantity in the square brackets, which we denote with $C_1(d)$, is
a slow-varying function of $d$ that is limited from below by 2 or so
and becomes a monotonically increasing function of $d$ for large
values of the latter variable. $C_1(d)$ tends to $2 \sqrt{2} \approx
2.8$ as $d \to \infty$. This is about 15\% greater than the value
reported in Ref.~\cite{10.2307/3182776}. At $d=1$, we have $\approx
2.1$, i.e., about 20\% below the (numerically) exact value obtained in
the simulation, see Fig.~\ref{llthFig}. Thus the closed-form
expression (\ref{taull}), while approximate, provides satisfactory
accuracy while preserving the essential scaling with $d$.

The value of the optimal rate itself
\begin{equation} \frac{\Delta t}{\tau_\text{opt}} = \frac{1}{d} \:
  \frac{(2 f_d)^{2/(d+2)}}{(2/d+1)^{d/2+1}} \equiv \frac{C_2(d)}{d}
  \xrightarrow[d \to \infty]{} \frac{4/e}{d}
\end{equation}
is evaluated near the stationary region of the actual function (whose
exact value we do not possess) and, thus, is less sensitive to the
details of the approximation. It gives $\tau_\text{opt}/\Delta t
\approx 2.4$ in 1D, in good agreement with Fig.~\ref{llthFig}, as well
as yielding the correct $d \to \infty$ scaling. The multiplicative
factor $4/e \approx 1.47$ is about 10\% off the estimate $2.38^2
\times 0.234 \approx 1.32$ reported in Ref.~\cite{10.2307/3182776}.

\begin{figure}
  \includegraphics[width=0.8
    \columnwidth]{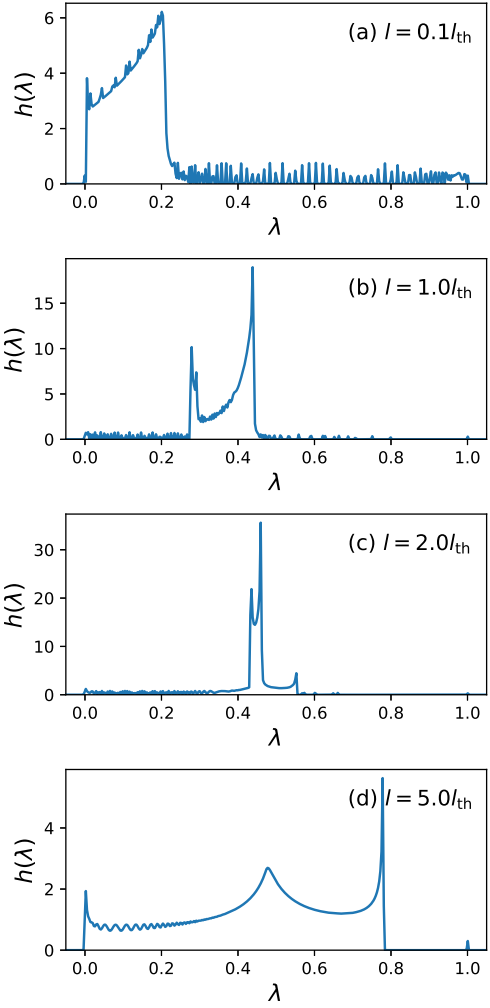}
  \caption{\label{gap2} The spectrum of the transition matrix $\pi$ of
    the Markov chain for four select values of the $l/\lth$
    ratio. Each energy level is represented by a narrow (width=0.001)
    Gaussian peak of unit area. $1001 \times 1001$.}
\end{figure}

Finally, in Fig.~\ref{gap2}, we show a coarse-grained spectrum of the
eigen-values $\lambda$ of the transition matrix $\bpi$ corresponding
to the parabolic energy function (\ref{EHO}), so as to complement
Fig.~\ref{gap}. The coarse-graining is performed by centering a narrow
Gaussian peak at each individual value of $\lambda$.  Note the
large-$\lambda$ end of the spectrum is difficult to see. For this end
of the spectrum, please consult Fig.~\ref{gap}.

%\newpage
%\bibliographystyle{naturemag}
%\bibliography{lowT}
\bibliography{/Users/vas/Documents/tex/ACP/lowT}
%\bibliography{/Users/vassiliylubchenko/Documents/tex/ACP/lowT}

\end{document}